\documentclass[twocolumn]{aastex631}
\usepackage{newtxtext,newtxmath}
\usepackage[T1]{fontenc}
\usepackage{graphicx}
\usepackage{amsmath,amssymb,mathtools}
\usepackage{bm}
\usepackage{booktabs}
\usepackage{multirow}
\usepackage{hyperref}
\usepackage{xfrac}
\usepackage{tikz}
\usepackage{color}

\newcommand{\be}{\begin{equation}}
	\newcommand{\ee}{\end{equation}}
\newcommand{\bea}{\begin{eqnarray}}
	\newcommand{\eea}{\end{eqnarray}}
\newcommand{\nn}{\nonumber}

\newcommand{\Mc}[1]{\mathcal{#1}}
\newcommand{\dif}{\mathrm{d}}

\newcommand{\upup}{\mathrel{\upharpoonleft\mkern-4mu\upharpoonright}}
\newcommand{\updown}{\mathrel{\upharpoonleft\mkern-4mu\downharpoonright}}

\def\sst#1{{\scriptscriptstyle #1}}
\def\0{{\sst{(0)}}}
\def\1{{\sst{(1)}}}
\def\2{{\sst{(2)}}}
\def\3{{\sst{(3)}}}
\def\4{{\sst{(4)}}}
\def\5{{\sst{(5)}}}
\def\6{{\sst{(6)}}}
\def\7{{\sst{(7)}}}
\def\8{{\sst{(8)}}}

\def\a{\alpha}
\def\b{\beta}
\def\c{\chi}
\def\d{\delta}
\def\D{\Delta}
\def\f{\frac}
\def\g{\gamma}

\def\G{\Gamma}

\def\l{\left}
\def\L{\Lambda}
\def\m{\mu} 
\def\n{\nu} 
\def\nn{\nonumber}

\def\p{\phi} 
 
\def\r{\right}
\def\s{\sigma}
\def\S{\Sigma}
\def\t{\theta}
\def\T{\Theta}
\def\k{\kappa}

\def\o{\omega}

\def\mP{{\mathcal P}}

\def\mB{{\mathcal B}}
\def\mL{{\mathcal L}}
\def\mT{{\mathcal T}}

\def\bag{\begin{aligned}}
	\def\eag{\end{aligned}}
\def\ba{\begin{array}}
	\def\ea{\end{array}}

\begin{document}

\title{Non-thermal Synchrotron Emission and Polarization Signatures during Black Hole Flux Eruptions}
\shorttitle{Nonthermal Synchrotron in MAD}
\shortauthors{Zhou et al.}

\correspondingauthor{Yehui Hou}
\email{yehuihou@sjtu.edu.cn}

\correspondingauthor{Minyong Guo}
\email{minyongguo@bnu.edu.cn}

\correspondingauthor{Bin Chen}
\email{chenbin1@nbu.edu.cn}

\author[0009-0001-0796-1547]{Fan Zhou}
\affiliation{School of physics and astronomy, Beijing Normal University, Beijing 100875, P. R. China}

\author[0009-0002-2360-2971]{Jiewei Huang}
\affiliation{School of Physics, Peking University, No.5 Yiheyuan Rd, Beijing 100871, P.R. China}

\author[0009-0007-4339-0570]{Yuehang Li}
\affiliation{School of physics and astronomy, Beijing Normal University, Beijing 100875, P. R. China}

\author[0000-0003-0869-4601]{Zhenyu Zhang}
\affiliation{Institute of Fundamental Physics and Quantum Technology, \& School of Physical Science and Technology, Ningbo University, Ningbo, Zhejiang 315211, China}

\author[0000-0002-9434-3930]{Yehui Hou}
\affiliation{Tsung-Dao Lee Institute, Shanghai Jiao-Tong University, Shanghai, 201210, P. R. China}

\author[0000-0001-5577-575X]{Minyong Guo}
\affiliation{School of physics and astronomy, Beijing Normal University, Beijing 100875, P. R. China}
\affiliation{Key Laboratory of Multiscale Spin Physics, Ministry of Education, Beijing 100875, P. R. China}

\author[0000-0003-4509-9705]{Bin Chen}
\affiliation{Institute of Fundamental Physics and Quantum Technology, \& School of Physical Science and Technology, Ningbo University, Ningbo, Zhejiang 315211, China}
\affiliation{School of Physics, Peking University, No.5 Yiheyuan Rd, Beijing 100871, P.R. China}

\begin{abstract}
In this work, we investigate synchrotron emission and the observational signatures of anisotropic non-thermal electrons during magnetic-flux eruptions in a magnetically arrested disk, using 3D GRMHD simulations. Non-thermal electrons are assumed to be accelerated from the thermal background through magnetic reconnection, with pitch-angle distributions modeled as beamed or loss-cone types, alongside an isotropic case for comparison. The results show that non-thermal emission can produce pronounced flux outbursts and localized brightening during eruptions, while the associated increase in optical depth can suppress the linear polarization fraction. Introducing pitch-angle anisotropy further reshapes the angular distribution of the intrinsic emissivity and modulates its contribution to various observable signatures.
Strong field-aligned beaming in the electron distribution suppresses non-thermal emission for near-axis observers, effectively driving the image morphology toward a purely thermal limit. In contrast, moderately anisotropic models remain effective at imprinting non-thermal electron signatures on both the total intensity and polarization structure. We further quantify how eruption-driven increases in absorption depth and enhanced Faraday effects reduce the linear polarization fraction and modify the azimuthal coherence of the polarization field. 
Overall, our results demonstrate that incorporating anisotropic non-thermal electrons is essential for a physically self-consistent interpretation of time-variable EHT polarimetric observations.
\end{abstract}

\section{\label{sec1}Introduction}

The Event Horizon Telescope (EHT) has obtained the highest-resolution images to date of the supermassive black holes M87* and Sgr A* \citep{M87_1, M87_2, M87_3, SgrA_1, SgrA_2, SgrA_3}, opening new opportunities to probe fundamental physics and plasma dynamics in the strong-gravity regime \citep{M87_5, M87_6, SgrA_4, SgrA_5, SgrA_6, EventHorizonTelescope:2020qrl, EventHorizonTelescope:2021dqv}. Polarization structures of horizon-scale synchrotron emission reveal ordered magnetic fields \citep{M87_7, SgrA_7}, whose geometry encodes signatures of the accretion state and even spacetime rotation \citep{M87_8, SgrA_8}. Black-hole accretion systems also show variability \citep{Zhao:2000yd}: multi‑wavelength observations of Sgr A* reveal flares with enhanced emission \citep{Genzel:2003as, Ghez:2003hb, Baganoff:2001kw, GRAVITY:2020xcu, SgrA_4} and localized bright spots \citep{Abuter:2018uum, GRAVITY:2020lpa, Wielgus:2022heh, GRAVITY:2023avo, Ripperda:2020bpz}; recent time-variable polarized images of M87* implies a non-quiescent flow \citep{EventHorizonTelescope:2025vum}. These findings necessitate a better understanding of the dynamical behavior of black-hole accretion flows and their millimeter-wavelength emission and imaging properties.

General relativistic magnetohydrodynamic (GRMHD) simulations are now the standard tool for modeling magnetized accretion flows around black holes  \citep{EventHorizonTelescope:2019pcy, Gammie_2003}. When coupled with general relativistic radiative transfer (GRRT) \citep{Broderick:2003fc, Shcherbakov:2010kh, Dexter:2016cdk, Moscibrodzka:2017lcu, Pu:2018ute, Younsi:2019iee, Bronzwaer:2020kle, Aimar:2023vcs, noble2007simulating}, they successfully reproduce key observational features \citep{M87_5, M87_6, M87_8, M87_9, SgrA_4, SgrA_5, SgrA_6, SgrA_8, Davelaar:2017kgb}. Various studies suggest that M87* favors a magnetically arrested disk (MAD), in which strong, ordered magnetic flux threads the black hole \citep{Narayan2003}. This configuration naturally accounts for the observed jet properties \citep{Blandford1977, McKinney2004, Tchekhovskoy:2011zx, CruzOsorio2022, Sheperd2012} and spiral polarization patterns \citep{M87_7,M87_8, SgrA_7, SgrA_8}. MADs also exhibit recurrent magnetic-flux eruptions: excess flux accumulated on the horizon is intermittently expelled and later re‑accreted \citep{igumenshchev2008magnetically,Tchekhovskoy:2011zx,McKinney2012}. 
These eruptions release magnetic energy, drive turbulence and reconnection, and efficiently heat electrons, boosting the emitted flux, as shown by high-resolution simulations \citep{Porth:2020txf, GRAVITY:2021hxs, Ripperda_2022, Chatterjee:2022mxg, Najafi-Ziyazi:2023oil}. 
MAD flux-eruption events have therefore been proposed as a mechanism for black-hole flares \citep{Dexter:2020cuv, Porth:2020txf, Ripperda:2021zpn, Scepi:2021xgs, Jia2023, Grigorian:2024rsn, Antonopoulou:2025tgs, Jiang:2025huk, Jiang:2024gtk}. 

Electrons can be accelerated out of the thermal pool to high energies during flux-eruption events, generating power-law tails \citep{Moscibrodzka:2013rta, Porth2017, Davelaar:2019jxr, Chael:2018aeq}. 
The non-thermal population can substantially alter the synchrotron emissivity, effective optical depth, and both the degree and morphology of polarizations \citep{Yang:2018zrd, Fromm:2021mqd, Galishnikova:2023ltq, Tsunetoe:2024uzh, Zhang:2024ddt}. 
Moreover, the acceleration mechanisms such as magnetic reconnection are intrinsically anisotropic, implying that the accelerated non-thermal electrons should also exhibit anisotropic distributions \citep{Ball_2018,Comisso_2019,comisso2022ion, Comisso_2023,Comisso:2024iyx}. 
Additional processes near the black hole, such as particle escape along open magnetic field lines and the rapid cooling of high‑energy synchrotron emitters, can further imprint anisotropic velocity structures \citep{Kunz:2014qha, Riquelme:2014rha}. 
Such anisotropies can influence synchrotron emissions and image morphology \citep{Galishnikova:2023ltq,Tsunetoe:2024uzh,Tsunetoe:2025crz,Lai_2025,Yang:2018zrd}. 
Taking into account of anisotropic, non-thermal electrons generated from the thermal pool is therefore essential for building physically self-consistent models of MAD flux-eruption events and for interpreting their observational signatures, including polarimetric images and the distinctions between different electron-acceleration mechanisms.

In this work, we perform 3D GRMHD simulations to generate a MAD state around a spinning black hole, and systematically investigate how non-thermal, anisotropic electrons affect the total flux, spatially resolved images, polarization fractions and patterns, during flux-eruption episodes. 
The electrons are modeled using the $R-\beta$ prescription that relates the electron temperature to the simulated ion temperature and plasma $\beta$ \citep{Monika2016}. Synchrotron emission profile is produced by a combined population of thermal and non-thermal power-law electrons, with the non-thermal component assumed to be energized primarily by magnetic reconnections \citep{Ball_2018}. For the non-thermal electrons, we further incorporate various beam-like or loss-cone anisotropies that modulate the local emissivity \citep{Lai_2025}. We then perform GRRT calculations to generate 230 GHz polarized images for different electron distributions, enabling a detailed assessment of how non-thermal populations and anisotropy influence the resulting observables.

The structure of this paper is as follows. 
In Sec.~\ref{sec:grmhd}, we describe the setup and result of GRMHD. 
In Sec.~\ref{sec:eDF}, we construct the electron distributions and the combined emission model.
In Sec.~\ref{sec:imagingresults}, we present the GRRT result and analyze the total flux, intensity maps, linear polarization degree and polarization pattern. 
We summarize the main findings in Sec.~\ref{sec:conclusion} and discuss the directions for future work.
In the following, lengths and times are expressed in units of $r_{\text{g}}=GM/c^2$ and $t_{\text{g}}=GM/c^3$, where $M$ denotes the central black hole mass; 
electromagnetic fields are given in Gaussian units.

\section{GRMHD Simulation}\label{sec:grmhd}

\subsection{Basic equations}\label{sec:conser}

We model the accreting plasma as a non-self-gravitating magneto-fluid evolving in a fixed black hole spacetime. Its total energy–momentum tensor can be decomposed into matter and electromagnetic parts, $T^{\m\n} = T^{\m\n}_{\rm m} + T^{\m\n}_{\rm EM}$, with 
\bea
\begin{aligned}
	&T^{\m\n}_{\rm m} = h \, u^{\m}u^{\n} + p \, g^{\m\n} \,, \\
	&T^{\m\n}_{\rm EM} = F^{\m\rho}F^\n_\rho-\frac{1}{4}g^{\m\n}F^{\alpha \beta}F_{\alpha \beta} \, ,
\end{aligned}
\eea
where $h = e + p$ is the comoving-frame ion enthalpy density, with $e$, $p$ the internal energy density and pressure; $u^{\m}$ is the bulk four-velocity, and $F^{\m\n}$ is the Faraday tensor.
In the ideal-MHD limit, the plasma resistivity is set to zero, implying a vanishing comoving electric field: $u_\m F^{\m\n}=0$.

The dynamical evolution is governed by the GRMHD equations, consisting of (i) local energy-momentum conservation, $\nabla_ \m T^{\m\n} = 0 $, (ii) particle number conservation, $\nabla_\m(\rho u^{\m}) = 0 $ with $\rho$ the rest-mass density, and (iii) the Bianchi identity, $\partial_{[\lambda} F_{\mu\nu]} = 0$.  
These equations are supplied with an ideal-gas equation of state. Under the adiabatic approximation, the energy density satisfies the polytropic relation $e = \rho + \left(\hat{\gamma} - 1\right)^{-1}p$, where $\hat{\gamma}$ is the adiabatic index, taking $\hat{\gamma} = 5/3$ for non‑relativistic and $\hat{\gamma} = 4/3$ for relativistic ions.

For convenience, we introduce ``pseudo-electromagnetic'' fields defined as $E^{\m} = F^{t\m}$, $B^{\m} =  -(* F)^{t\m} $.
Their relation to the comoving-frame field $b^{\m} = -u_{\nu}(* F)^{\mu\nu}$ is given by $\{E^{\m}, B^{\m}\} = \{- \epsilon^{t\m\a\b}b_{\a}u_{\b}\,,b^{[\m}u^{t]} \}$, $\{ b^{t},b^i \} = \{ B^{\m}u_{\m} \,,\left( B^i+ b^t u^i \right)/u^t \}$, where $i$ denotes spatial indices. The pseudo-fields differ from the fields measured by spacetime normal observers only by a lapse function \citep{komissarov2004electrodynamics}.

\subsection{\label{sec2}Numerical setup}

We perform three‑dimensional GRMHD simulations of Kerr black hole accretion using the code \textbf{BHAC} \citep{Porth2017}. 
The equations are solved in spherical modified Kerr–Schild (MKS) coordinates \citep{McKinney2004}, with a spin parameter $a=0.9375$. 
The employment of static mesh refinement (SMR) yields an effective resolution of $N_r \times N_{\theta} \times N_{\phi} = 384 \times 192 \times 256$.  
Owing to the scale invariance of the GRMHD equations, the maximum initial mass density is normalized to $\rho_\text{max}=1$ during the simulation. 
To avoid unphysical vacuum states where the fluid description ceases to apply, numerical floor values are imposed: if $\rho \leq \rho_{\text{fl}} = 10^{-5} r^{-3/2}$ we set $\rho = \rho_{\text{fl}}$, if $p \leq p_{\text{fl}} = 1/3 \times 10^{-7} r^{-5/2}$ we set $p = p_{\text{fl}}$.

The initial equilibrium torus corresponds to the Fishbone–Moncrief solution \citep{Fishbone1976}, with inner radius $r_{\text{in}}=20$, pressure maximum at $r_{\text{max}}=40$, and adiabatic index $\hat{\gamma}=5/3$.
A poloidal magnetic field is seeded via the vector potential
\begin{equation}
	A_{\phi} = A_0 \left(\f{\rho}{\rho_\text{max}} - 0.01\right)\left( \f{r \sin \theta}{r_{\text{in}}} \right)^3 e^{-r/400 r_\text{g}}\,,
\end{equation}
with all other components vanishing. The normalization constant $A_0$ is chosen such that the minimum ratio of gas to magnetic pressure satisfies $\left(p/p_{\text{b}}\right)_{\text{min}} = 100$, where $p_{\text{b}}=b^{\mu}b_{\mu}/2$ is the magnetic pressure. This initialization ensures rapid growth of turbulence and saturation of the magnetorotational instability (MRI).
We further introduce the dimensionless magnetization parameter as $\sigma_\text{M} = b^2/\rho$. 
Extremely high values of $\sigma_\text{M}$ can lead to numerical instabilities and unreliable results. To avoid this, we impose a numerical ceiling $\sigma_\text{M,floor} = 50$, replenishing plasma in any cells where $\sigma_\text{M,floor}$ exceeds this value. For the subsequent analysis, a stricter threshold $\sigma_\text{M,cut} = 20$ is adopted, since the MHD approximation itself is expected to break down in strongly magnetized regions \citep{kulsrud2005plasma}. Zones with  $\sigma_\text{M} > \sigma_\text{M,cut}$ are therefore excluded from the analysis.

\subsection{MAD evolution}\label{sec:evolution}

For clarity of presentation, physical quantities from the simulation are plotted in Cartesian coordinates associated with the spherical Kerr-Schild (KS) system. 
The $x-z$ plane corresponds to $\phi = 0$ and $\pi$, while the $x-y$ plane corresponds to the equatorial plane at $\theta = \pi/2$.
The accretion state can be characterized in terms of the mass accretion rate $\dot{M}$ and the horizon-threading magnetic flux $\Phi_\text{EH}$, defined as \citep{Tchekhovskoy:2011zx}
\bea
\begin{aligned}
	&\dot{M} = \int_{r_\text{h}} \rho u^r \sqrt{-g} \dif \theta \dif \phi\,, \\ 
	&\Phi_\text{EH} = \frac{1}{2} \int_{r_\text{h}} |B^r| \sqrt{\gamma} \dif \theta \dif \phi\,,
\end{aligned}
\eea
where $g$ denotes the determinant of the Kerr metric and $\gamma$ the determinant of its spatial part; $r_\text{h} = M + \sqrt{M^2- a^2}$ is the horizon radius.
We further introduce a dimensionless MAD parameter $\phi_\text{EH} = \Phi_\text{EH} / \sqrt{\dot{M}}$, which provides a convenient measure of flux accumulation. An accretion flow reaches the MAD state once $\phi_\text{EH} \simeq 15$ \citep{Tchekhovskoy:2011zx}.

\begin{figure}[htbp]
\centering
\hspace*{-1cm}
	\includegraphics[width=0.57\textwidth]{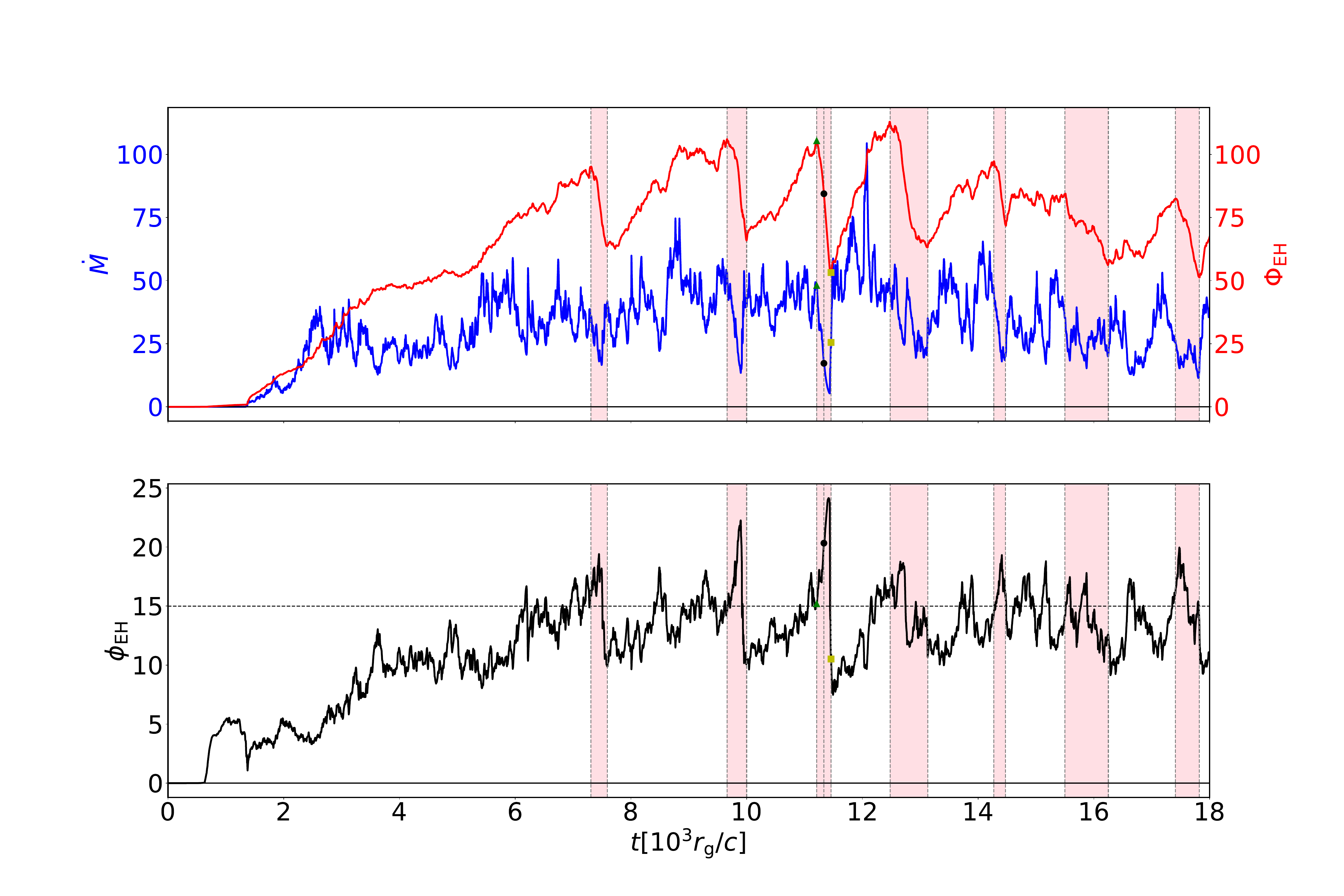}
	\caption{Time evolutions of accretion rate, magnetic flux, and the MAD parameter. 
The horizontal black solid lines denote the zero level, while in the lower panel the horizontal black dashed line marks $\phi_{\text{EH}} = 15$. We identify flux-eruption events as the pink bands, where $\Phi_\text{EH}$ drops steeply from a local maximum to a subsequent local minimum. The green triangles, black dots, and yellow squares denote the pre-eruption, peak, and post-eruption phases of the third flux eruption event, respectively.}
	\label{fig:mdot}
\end{figure}

The evolutions of the accretion rate and magnetic flux are shown in Fig.~\ref{fig:mdot}. Early on, MRI-driven angular‑momentum transport causes the initially equilibrium torus to lose angular momentum and transition into an inflow. Magnetic field lines are advected inward with the plasma and accumulate near the horizon, producing a steady rise in $\Phi_\text{EH}$. 
During this phase the flux remains unsaturated, and the system stays in the standard and normal evolution (SANE) state \citep{Narayan:2012yp, Sadowski:2013joa, M87_5}. 
Around $t \approx 6000 \,t_\text{g}$, $\phi_\text{EH}$ reaches 15, indicating saturation of the poloidal flux near the black hole. The resulting magnetic pressure impedes and partially disrupts the inflow, marking the onset of the MAD state. The flux then undergoes repeated cycles of growth and release. The sharp drops highlighted by the pink bands in Fig.~\ref{fig:mdot} correspond to flux‑eruption events \citep{igumenshchev2008magnetically, Tchekhovskoy:2011zx, Chatterjee:2022mxg}, 
which occur when the horizon-threading flux becomes oversaturated and is rapidly expelled along with matter. After each eruption, the system relaxes to a sub‑saturated state, enabling the next cycle of flux buildup and release.

\begin{figure*}[htbp]
	\centering
	\includegraphics[width=0.85\textwidth]{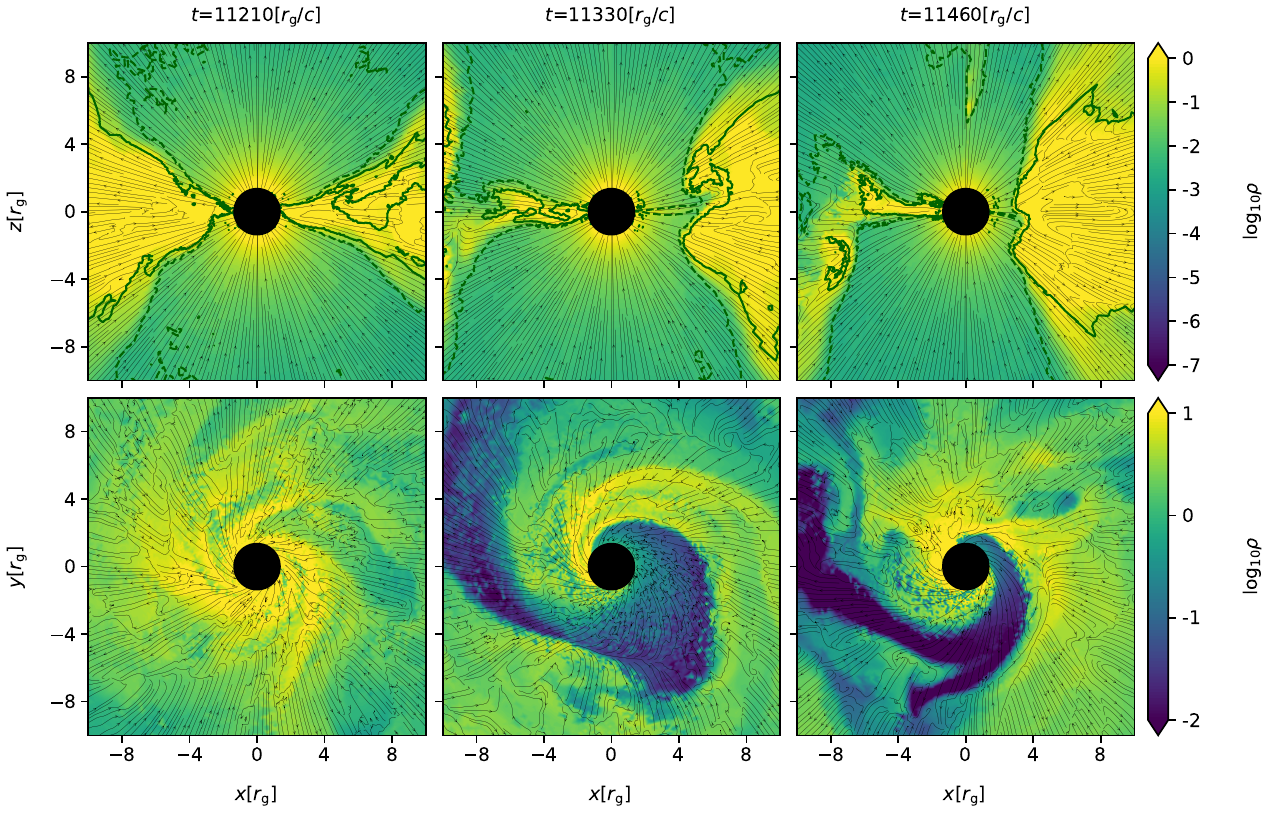}
	\caption{Density profiles in the $x-z$ plane (\textbf{top}) and $x-y$ plane (\textbf{bottom}) at $t = 11210 \,t_\text{g}$, $t = 11330 \,t_\text{g}$ and $t = 11460 \,t_\text{g}$. The dark green solid contour represents $Be = 1.05$, the dark green dashed contour indicates the magnetization $\sigma_\text{M} = 20$, and the black arrows depict the magnetic field lines (the same below).}
	\label{fig:rho}
\end{figure*}

Fig.~\ref{fig:rho} shows the ion density immediately before, during, and after the third eruption episode ($11210 \,t_\text{g} \sim 11460 \,t_\text{g}$).
In the $x-z$ plane, we plot contours of the Bernoulli parameter $Be = -hu_t / \rho = 1.05$ (solid curves) and of $\sigma_\text{M} = \sigma_\text{M,cut}$ (dashed curves), which partition the flow into three zones: the disk, with $Be < 1.05$, the jet sheath, with $Be > 1.05$, $\sigma_\text{M} < \sigma_\text{M,cut}$, and the jet spine, with $\sigma_\text{M} > \sigma_\text{M,cut}$ \citep{McKinney2012, CruzOsorio2022}, where the plasma becomes sufficiently tenuous to require force-free treatment \citep{Chandra:2015iza, Chael:2024gvx}.
Prior to the eruption, the density exhibits a standard, approximately axisymmetric structure. During the eruption, turbulent features develop near the midplane, and the sheath boundaries become strongly distorted. Both the density and magnetic-field morphology indicate that the eruption originates near the equatorial region \citep{Ripperda_2022, Jia2023}.
At peak eruption ($t = 11330 \,t_\text{g}$), a low-density, highly magnetized region with a swirling pattern appears in the $x-y$ plane due to magnetic-pressure release.
In the post-eruption stage ($t = 11460 \,t_\text{g}$), the system shows partial recovery, but the outflow retains an asymmetry. 
The eruption phase is likewise characterized by outward flows, increased magnetization, stronger toroidal magnetic field, and enhanced rotation rate, as illustrated in Appendix \ref{app:grmhd}.

\section{Modeling the Electrons}\label{sec:eDF}

\subsection{Basic setup}\label{sec:eDFbasis}

In radiatively inefficient accretion flows, protons and electrons attain different temperatures because Coulomb coupling is weak in the low-density plasma \citep{M87_5, M87_6, Dihingia:2022aoc}. The proton temperature relates to the gas pressure via $\rho \, k_\text{B} T_\text{p} =  m_\text{p}\, p_\text{g}$. The electron temperature is estimated using the empirical $R-\beta$ model,
\begin{equation}\label{eq:Rbeta}
	\frac{T_\text{p}}{T_\text{e}} = R_\text{h} \frac{\beta^2}{1+\beta^2} + R_\text{l} \frac{1}{1+\beta^2}\,,
\end{equation}
where $\beta=p_{\text{g}}/p_{\text{b}}$ is the plasma-$\beta$ parameter. The constants $R_\text{h}$ and $R_\text{l}$ set the characteristic proton–electron temperature ratios in strongly and weakly magnetized regions, respectively, thereby modeling the $\beta$-dependent coupling \citep{Monika2016}. Synthetic images produced with this prescription match those obtained from magnetic-reconnection–based electron-heating models \citep{Mizuno:2021esc}.

The electron energy spectrum and radiative properties are set by the phase‑space distribution measured in the fluid comoving frame. For a gyrotropic distribution—i.e., isotropic in the plane perpendicular to the magnetic field—the electron distribution function (eDF) is
\bea\label{generaleDF}
\frac{\dif n_\text{e}}{\dif \gamma \dif \Omega} = n_\text{e}\, F(\gamma) \, G(\alpha) \,,
\eea
where $n_\text{e}$ is the local number density, $\gamma$ the Lorentz factor, and $\alpha$ is the pitch angle between the electron momentum and the local magnetic field. 
The functions $F(\gamma),G(\alpha)$ are normalized such that $\int^{\infty}_{1}F(\gamma)\dif\gamma = 2\pi \int^{\pi}_{0}G(\alpha)\dif\cos{\alpha} = 1$. 
Note that Eq.~\eqref{generaleDF} assumes a separable form; in general, $\gamma, \alpha$ are coupled \citep{Galishnikova:2023ltq}, and this form was also used in \citet{Tsunetoe:2025crz}. For relativistic electrons, however, an expansion in $\gamma^{-1}$ always yields a separable leading-order term.

Electrons in quiescent accretion disks are often assumed to thermalize into an isotropic Maxwell-Jüttner (MJ) distribution \citep{juttner1911maxwellsche}. The mean energy per particle is $ m_{\text{e}} f\left(\Theta_\text{e}\right)\Theta_\text{e}$, where $\Theta_\text{e} = k_BT_{\text{e}}/m_{\text{e}}$ is the dimensionless temperature and $f\left(\Theta_\text{e}\right) = \left( 6 + 15 \Theta_\text{e} \right) \left( 4 + 5 \Theta_\text{e} \right)^{-1}$ is the relativistic correction factor \citep{gammie1998advection}. When the electron relaxation time is short compared to the dynamical timescale, the MJ form is adequate.
In weakly collisional flows, however, processes such as reconnection and shocks can accelerate a fraction of electrons, generating a high-energy tail \citep{Moscibrodzka:2013rta, Davelaar:2019jxr, quataert2002magnetorotational}. Over a narrow frequency range, the non-thermal electrons can be adequately represented by a single power-law distribution:
\begin{equation}
	F_p(\gamma) = \f{p-1}{\gamma_{\rm min}^{1-p}-\gamma_{\rm max}^{1-p}}\, \gamma^{-p}  \,,  \quad \text{for} \,\, \gamma_{\rm min} \leq \gamma \leq \gamma_{\rm max} \,,
\end{equation}
where $p$ is the spectral index. The low-energy cutoff $\gamma_{\rm min}$ is typically set at the peak of the MJ distribution, $\gamma_{\rm min} = 1 + f\left( \Theta_\text{e} \right) \Theta_\text{e}$, while the high-energy cutoff $\gamma_{\rm max}\sim 10^5$ is determined by microphysical processes \citep{Melzani2014}. Because the results are insensitive to the precise value of $\gamma_{\rm max}$, it is often taken formally to infinity. The mean electron energy is then $m_{\text{e}} \int^{\gamma_{\text{max}}}_{\gamma_{\text{min}}}\left(\gamma-1\right) F_p(\gamma) d \gamma \approx m_{\text{e}} \left( \f{p-1}{p-2}\gamma_{\text{min}} - 1 \right)$.
For reconnection‑driven heating, particle-in-cell (PIC) simulations provide an empirical calibration of $p$ in terms of $\beta$ and $\sigma_\text{M}$ \citep{Ball_2018}:
\begin{equation}\label{eq:PICp}
	p(\beta, \sigma_\text{M}) = 1.8 + 0.7 \sigma_\text{M}^{-0.5} + 3.7 \sigma_\text{M}^{-0.19} \tanh(23.4  \sigma_\text{M}^{0.26} \beta)   \,. 
\end{equation}
The index decreases with increasing magnetization; for example, $p \approx 5.54$ at $\sigma_M = 2$, and $p \approx 4.84$ at $\sigma_M = 5$. Its dependence on $\b$ is rather weak as long as $\b \gtrsim 0.1$. 
Because the non-thermal population considered here is generated primarily in eruptive events, turbulent-reconnection heating dominates, and Eq.~\eqref{eq:PICp} provides an appropriate characterization.

\subsection{Angular dependence}\label{sec:anisotropiceDF}

In dynamically strong magnetic fields, weakly collisional electrons readily develop anisotropies. Although the eDF remains gyrotropic, its parallel and perpendicular components can differ substantially \citep{kulsrud1983mhd}. 
Studies have shown that such anisotropy can modify the synchrotron spectrum \citep{Yang:2018zrd, Lai_2025}. It may also imprint observable signatures in mm-band images. 
To examine anisotropy in a controlled way, we adopt two Gaussian-type prescriptions \citep{Lai_2025}. The first is a beam-like distribution, naturally produced by reconnection in relativistic, magnetically dominated plasmas \citep{Comisso_2019, comisso2022ion, Comisso_2023}:
\bea\label{beam1}
&& G_{b}(\alpha) = \frac{1}{X} \exp\left(-\frac{(\cos\alpha - \cos\alpha_0)^2}{2\sigma^2}\right)\,, \\
&& X = \sqrt{2\pi^3 \sigma^2}\left[\text{erf}(t_2) - \text{erf}(t_1)\right] \,, \nn \\
&& t_2 = \frac{1}{\sqrt{2\sigma^2}}(1 - \cos\alpha_0) \,, \quad t_1 = -\frac{1}{\sqrt{2\sigma^2}}(1 + \cos\alpha_0) \,, \nn
\eea
where $\alpha_0$, $\sigma$ set the beam center and width, respectively. The values $\a_0 = 0$ and $\pi$ correspond to beams aligned parallel or antiparallel to the magnetic field, respectively.
The second distribution is a loss-cone form, representing depleted electron populations along a given direction. Small-pitch-angle electrons on open magnetic field lines could be captured by the horizon, leading to their depletion, while magnetic mirroring suppresses particles moving nearly parallel to the field \citep{Kunz:2014qha, Riquelme:2014rha}. These effects generate a loss-cone-like eDF, modeled as
\bea\label{loss1}
G_{l}(\a) = \f{1}{4\pi - X} \left[1- \exp\left(-\frac{(\cos\a - \cos\a_0)^2}{2\sigma^2}\right) \right]\,, 
\eea
where $\a_0 = 0$ and $\pi$ yields loss cones oriented parallel or antiparallel to the magnetic field. Radiative cooling can also preferentially deplete large-pitch‑angle, high-energy electrons, effectively producing a loss cone centered at $\a_0 = \pi/2$, captured by the same functional form in Eq.~\eqref{loss1}.

Note that the electron distributions in Eq.~\eqref{beam1} and Eq.~\eqref{loss1} do not possess $Z_2$ symmetry with respect to the plane normal to the magnetic field (except when $\alpha_0 = \pi/2$), implying direction-dependent acceleration. The specific preferred direction, however, is not known. In the following, we also consider a more conservative $Z_2$-symmetrized construction in which the eDF is defined as $\tilde{G}(\a) = \left[ G(\a) +  G(\pi-\a)\right]/2$. We refer to these symmetrized cases as the bi-beam and bi-loss-cone eDFs.

Because the acceleration mechanisms are not precisely known, the fractional contribution of each anisotropic component to the total electron population is uncertain. We therefore treat $\alpha_0,\sigma$ as free parameters and analyze the synchrotron emission from each eDF in Eq.~\eqref{beam1}, Eq.~\eqref{loss1} and their symmetrized forms separately, without attempting to model any kinetically determined mixtures. A more rigorous treatment is left for future work.

\subsection{Anisotropic emissivity}\label{sec:emissivityplot}

\begin{figure}[htbp]
	\centering
\hspace*{-1.5mm}
	\includegraphics[width=0.48\textwidth]{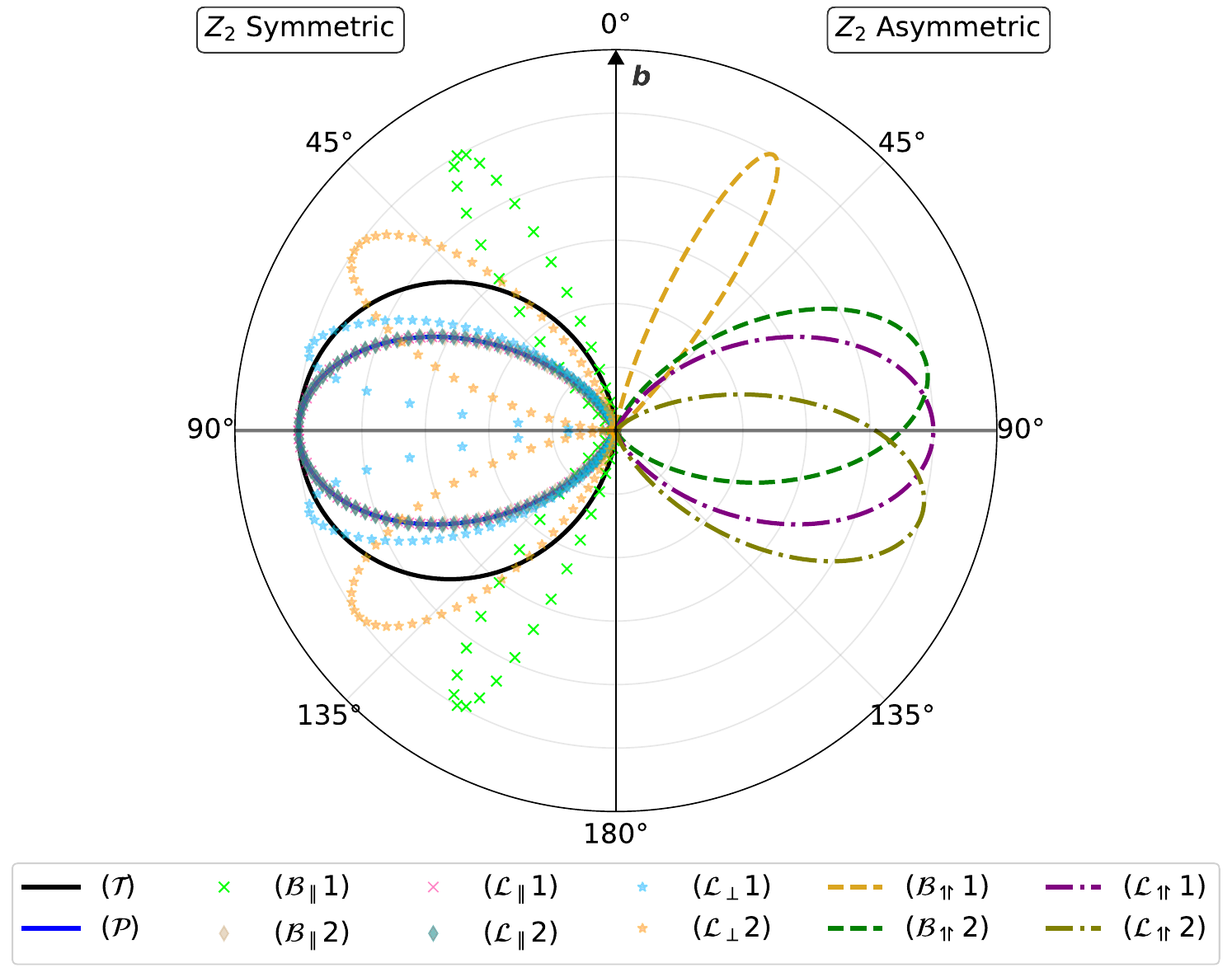}
	\caption{Angular dependence of synchrotron emissivities for eDFs with with (\textbf{left}) and without (\textbf{right}) the $Z_2$ symmetry, shown as functions of $\a$ within the fluid comoving frame. The axis $\a = 0^{\circ}$ denotes the direction of the local magnetic field $b$.}
	\label{fig:emissivity}
\end{figure}

Prior to the imaging analysis, it is useful to examine how the local emissivities behave for different eDFs. Fig.~\ref{fig:emissivity} shows the synchrotron emissivities for several fiducial models as functions of the pitch angle $\a_k$. 
In our notation, $\mathcal{B}$ and $\mathcal{L}$ denote the beam and loss‑cone eDFs, respectively; the subscripts $\upup$, $\updown$ and $\parallel$ indicate the cases $\a_0 = 0, \pi$ and the $Z_2$ symmetrized configuration, while the subscript $\perp$ indicates $\a_0 = \pi/2$; the labels ``1'', ``2'' refer to $\sigma = 10^{-1}$ and $1$, respectively.
The parameter choices are summarized in Table.~\ref{tab:rescale}. 
We have also examined a much narrower beam width with $\sigma=0.01$. 
Under a near-axis viewing geometry and parameters appropriate for M87*, such extreme collimation directs the nonthermal emission largely away from the line of sight. As a result, the observed flux is dominated by the residual thermal background, rendering the $\sigma=0.01$ beamed models effectively indistinguishable from the purely thermal case (see also Sec. \ref{sec:effectofaniso}). In contrast, the corresponding loss-cone models asymptotically approach the isotropic hybrid limit, as is evident from Eq. \eqref{loss1}.

The angular dependence of the emissivity reflects the combined effects of the single‑electron synchrotron pattern and the eDF. A relativistic electron emits strongly along its direction of motion (the headlight effect), and its emissivity is written as $J^e_\nu \approx P^e_\nu(\gamma, \a) \, \delta(\Omega-\Omega_k)$, where $P^e_\nu$ is the total radiated power. 
Integrating $J^e_\nu$ over the distribution function Eq.~\eqref{generaleDF} yields the synchrotron emissivity 
\bea\label{ultraemission}
j_{\nu}(\a_k) =  n_e G(\a_k) \int^{\infty}_{1} \dif \gamma F(\gamma) P^{e}_\n(\gamma, \a_k) \,.
\eea
As $\a_k \to 0$ or $\pi$, $P^e_\nu$ decreases sharply, producing the intrinsic synchrotron anisotropy already visible in the isotropic-eDF results in  Fig.~\ref{fig:emissivity}.
Anisotropy in the eDF further modulates the structure of $j_{\nu}$. 
For the bi-beam model $\mB_{\parallel}1$, the electrons aligned with the magnetic field suppress emission near the normal plane and generate a double-cone enhancement around $\alpha_k \approx 30^{\circ}$. 
For $\mB_{\parallel}2$, the weaker anisotropy is largely overwhelmed by the intrinsic synchrotron pattern, resulting in an emissivity similar to that of $\mP$.

For $Z_2$-asymmetric cases, $j_\nu$ for $\mB_{\upup}1$ resembles the one for $\mB_{\parallel}1$ in the northern hemisphere but vanishes in the southern one.
The emission from $\mB_{\upup}2$ is more moderate, with its maximum shifted toward the normal plane.
For the loss‑cone models, the distribution for $\mL_{\upup}1$ (and its symmetrized case $\mL_{\parallel}1$) are nearly isotropic, as implied by Eq.~\eqref{loss1}, yielding emissivities close to $\mP$;
for $\mL_{\upup}2$, the emissivity peak shifts slightly toward the southern hemisphere.
In the bi‑loss‑cone case $\mL_{\parallel}2$, symmetrization smooths both polar cones, again producing a pattern similar to $\mP$.
For $\mL_{\perp}1$ and $\mL_{\perp}2$, the loss cone suppresses emission near the normal plane and generates a weaker double‑cone structure reminiscent of $\mB_{\parallel}1$ and $\mB_{\upup}1$.
In what follows, we omit eDFs whose emissivities are degenerate with other cases. We therefore focus on $\mT$, $\mP$, $\mB_{\parallel}1$, $\mB_{\upup}1$, $\mB_{\upup}2$, $\mB_{\updown}1$, $\mB_{\updown}2$, $\mL_{\upup}1$, $\mL_{\updown}1$, and $\mL_{\perp}1$, $\mL_{\perp}2$.

\subsection{Combined emission model}\label{sec:emissionmodel}

Since non-thermal electrons are excited from the thermal background, both thermal and non-thermal synchrotron emission must be included when constructing the emissivity of the accretion flow \citep{Fromm:2021mqd, SgrA_5}. The resulting eDF consists of an isotropic thermal component together with a possibly anisotropic non-thermal component, taking the form
\bea\label{combeDF}
\frac{\dif n_{\text{e}}}{\dif \gamma \dif \Omega} = n_\text{th} \f{F_{\text{MJ}}(\gamma)}{4\pi} + n_\text{nth} F_p(\gamma) G(\alpha) \,,
\eea
where the subscripts “th” and “nth” denote thermal and non‑thermal contributions; $F_{\text{MJ}}(\gamma)$ denotes the Maxwell–Jüttner distribution. 
The ratio of non-thermal electron energies to the total energies can be read from the PIC simulation \citep{Ball_2018}, $\epsilon(\beta, \sigma_\text{M}) = A_{\epsilon} + B_{\epsilon} \tanh{\left(C_{\epsilon}\beta\right)}$,
where $A_{\epsilon} = 1 - \left( 4.2 \sigma_\text{M}^{0.55}+1 \right)^{-1}$, $B_{\epsilon} = 0.64 \sigma_\text{M}^{0.07}$, and $C_{\epsilon} = -68 \sigma_\text{M}^{0.13}$ . 
Thus, the number-density ratio can be derived as
\bea\label{ratioR}
\mathrm{R} = \f{n_{\text{nth}}}{n_{\text{th}}} = \f{f\left(\Theta_\text{e}\right)\Theta_\text{e}}{\left[\left(\f{p-1}{p-2}\right)\gamma_{\text{min}} - 1 \right]}\f{\epsilon}{1 - \epsilon} \,,
\eea
where we have sent the upper bound $\gamma_{\text{max}}$ to infinity.
Combining Eq.~\eqref{ratioR} with the neutrality condition $n_{\text{e}} = n_{\text{p}} = \rho/m_{\text{p}}$, the eDF is fully determined. Thus, the synchrotron emissivity ($j_\nu$), absorptivity ($\alpha_\nu$), and Faraday rotation coefficients ($\rho_\nu$) from thermal and non-thermal electrons are computed as the weighted sums:
\begin{equation}\label{combemission}
	c_\nu = \f{1}{1 + \mathrm{R}} c_{\text{th},\nu} + \f{\mathrm{R}}{1 + \mathrm{R}} c_{\text{nth},\nu}\,,
\end{equation}
where $c_\nu = \left\{ j_\nu, \alpha_\nu, \rho_\nu \right\}$.   
Eq.~\eqref{combemission} enhances the contribution of non-thermal particles in regions of high magnetization, consistent with simulated results.
For calculating the synthetic images, we use analytical fits for $\left\{ j_\nu, \alpha_\nu, \rho_\nu \right\}$ corresponding to thermal and power-law eDFs \citep{Dexter:2016cdk, Marszewski:2021fkr}. 
With anisotropy Eqs.~\eqref{beam1}–\eqref{loss1}, $j_\nu, \alpha_\nu$ retain closed forms in the ultra-relativistic limit \citep{leung2011numerical, pandya2016polarized, Galishnikova:2023ltq}, as in Eq.~\eqref{ultraemission}, so the anisotropic factor enters simply into $G(\alpha_{k})$. 
Faraday rotation coefficients, however, are not analytically tractable for anisotropic eDFs, and we therefore resort to the isotropic-eDF expressions for $\rho_V$. 

\begin{table*}[ht]
	\centering

	\small
	\caption{Accretion rates for the fiducial eDF models, represented by calligraphic symbols for convenience.}
	\label{tab:rescale}
	\setlength{\tabcolsep}{6pt}
	\renewcommand{\arraystretch}{1.1}

\begin{minipage}{0.8\textwidth}
	\centering
	
	\begin{tabular}{lcccccccc}
		\hline
		\textbf{Model} &
		$\mT$ &
		$\mP$ &
		$\mB_{\upup}1$ &
		$\mB_{\upup}2$ &
		$\mB_{\updown}1$ &
		$\mB_{\updown}2$ &
		$\mB_{\parallel}1$ \\
		\hline
		\hline
		eDF &
		thermal &
		thermal + power-law &
		\multicolumn{4}{c}{thermal + beam} & thermal + bi-beam  \\
		\hline
		$\alpha_0$ (rad) & / & / & 0 & 0 & $\pi$ & $\pi$ & 0 or $\pi$ \\
		\hline
		$\sigma$ & / & / & $10^{-1}$ & $1$ & $10^{-1}$ & $1$ & $10^{-1}$ \\
		\hline
		$\dot{M}$ ($10^{-4}\,M_\odot\,\mathrm{yr^{-1}}$) &
		5.5 & 4.5 & 5.0 & 4.4 & 5.6 & 4.6 & 5.2 \\
		\hline
	\end{tabular}

	\vspace{4pt}

\end{minipage}

\begin{minipage}{0.8\textwidth}
	\centering
	\hfill

	\begin{tabular}{lcc@{\hspace{10mm}}cccccc}
		\hline
		\textbf{Model} &
		$\mL_\perp1$ &
		$\mL_\perp2$ &
		$\mL_{\upup}1$ &
		$\mL_{\upup}2$ &
		$\mL_{\updown}1$ &
		$\mL_{\updown}2$ &
		$\mL_{\parallel}1$ \\
		\hline
		\hline
		eDF &
		\multicolumn{6}{c}{thermal + loss-cone } & thermal + bi-loss-cone  \\
		\hline
		$\alpha_0$ (rad) & $\pi/2$ & $\pi/2$ & 0 & 0 & $\pi$ & $\pi$ & 0 or $\pi$ \\
		\hline
		$\sigma$ & $10^{-1}$ & $1$ & $10^{-1}$ & $1$ & $10^{-1}$ & $1$ & $10^{-1}$ \\
		\hline
		$\dot{M}$ ($10^{-4}\,M_\odot\,\mathrm{yr^{-1}}$) &
		4.6 & 4.8 & 4.5 & 4.8 & 4.4 & 4.3 & 4.5 \\
		\hline
	\end{tabular}
	\hfill
\end{minipage}
\end{table*}

\section{Imaging Results}\label{sec:imagingresults}

The emission profile of the accretion flow is specified by combining the GRMHD output in Sec.~\ref{sec:grmhd} with the eDFs introduced in Sec.~\ref{sec:eDF}, where we fix $R_\text{l} = 10$ and $R_\text{h} = 100$, which falls within the parameter space allowed by the EHT polarimetric constraints \citep{M87_8}. 
To generate synthetic images, we perform GRRT calculations using \textbf{Coport-2.0}, which is an updated version of \textbf{Coport} \citep{Huang2024}. This code solves polarized transfer in a covariant framework and incorporates both gravitational and plasma effects. It reads the AMR/SMR grids from \textbf{BHAC} and interpolates the simulation data along geodesics to compute the Stokes intensities (see Appendix.~\ref{app:coport} for details). 
The images are ray-traced at 230 GHz with a resolution of $512 \times 512$ pixels on an observer's image plane, located at $\theta_\text{o} = 17^\circ$ and $\phi_\text{o} = 180^\circ$. \footnote{For an axisymmetric emission profile, the results at $17^{\circ}$ and $163^{\circ}$ differ only by a reflection symmetry: the intensity and polarization are mirrored across the horizontal axis. In a time-variable flow, minor differences between northern and southern views may appear, but they merely reflect asymmetries about the equatorial plane.} 
We adopt the fast-light approximation \citep{moscibrodzka2009radiative, Gold:2016hld}, treating each GRMHD snapshot as an instantaneous emissivity field. 
To enable a more direct comparison with current EHT observations, we further convolve the synthetic Stokes images with a circular Gaussian beam with FWHM $=17\,\upmu$as and recompute the image-integrated observables. The results are presented both before and after convolution for straightforward comparison.

\subsection{Total flux}

The dimensionless GRMHD result should be rescaled to match the case of M87* \citep{M87_4}, with the scaling determined by requiring the averaged total flux within a quiescent period $t \in [10800 \,t_\text{g}, 11000 \,t_\text{g}]$ to be $\approx 0.66\,\text{Jy}$ at 230 GHz. 
Based on this procedure, we obtain accretion rates for different eDF models, which are summarized in Table~\ref{tab:rescale}. Note that we only focus on the third eruption event, as the imaging signatures from different events are qualitatively similar (see Fig.~\ref{fig:flux_diffevent} of Appendix.~\ref{app:imaging}). 
In Fig.~\ref{fig:flux}, the fluxes before and after convolution are overplotted. Their overlap reflects the fact that convolution with a normalized Gaussian kernel preserves the total flux density. The temporal evolution shown here is therefore intrinsic to the models.


\subsubsection{Effect of emission anisotropy}\label{sec:effectofaniso}

In the purely thermal case $\mT$, the required accretion rate is relativistically higher, whereas for $\mP$ it is lower. This difference arises because the observing frequency lies above the thermal synchrotron peak ($\nu_{\text{peak}} \sim 5\Theta_{\text{e}}^2 \nu_{\text{B}}$, where $\nu_{\text{B}} = eB/m_{\text{e}}$ is the cyclotron frequency) leading to suppressed thermal emission. In contrast, the non-thermal high-energy tail radiates efficiently near 230 GHz and remains bright, consistent with previous studies. Consequently, when non-thermal electrons are included, a lower accretion rate and density are sufficient to reproduce the fixed observed flux.

For the anisotropic models, $\mB_{\upup}2$, $\mB_{\updown}2$, $\mL_{\perp}1$, $\mL_{\upup}1$, $\mL_{\updown}1$, and $\mL_{\parallel}1$ have accretion rates similar to $\mP$, owing to their weak anisotropy. 
In contrast, $\mB_{\upup}1$, $\mB_{\updown}1$,  $\mB_{\parallel}1$, $\mL_{\upup}2$,and $\mL_{\perp}2$ 
show higher accretion rates and thus weaker intrinsic emission. 
This arises largely due to a mismatch between the preferred electron beaming directions (the peak of $G(\a)$) and the dominant synchrotron emission directions (the peak of $P^e_{\nu}(\a)$), which significantly suppresses the intrinsic emissivity.

Moreover, the flux of $\mB_{\upup}1$ is clearly higher than that of $\mB_{\updown}1$, whereas $\mL_{\upup}2$ is fainter than $\mL_{\updown}2$.
In the comoving frame, the peak emission of $\mB_{\upup}1$ and $\mL_{\updown}2$ is directed toward the northern side of the normal plane, while that of $\mB_{\updown}1$ and $\mL_{\upup}2$ points southward (Fig.~\ref{fig:emissivity}). 
For a nearly face-on observer, photons are preferentially emitted toward the northern hemisphere with pitch angles below $90^{\circ}$, owing to the split-monopole-like magnetic field in MAD  (Fig.~\ref{fig:rho}). 
As a result, $\mB_{\upup}1$ and $\mL_{\updown}2$ gain intrinsic emissivity, whereas part of the emissions from $\mB_{\updown}1$ and $\mL_{\upup}2$ is missed, reducing their observed flux. The situation reverses for an observer located near the south polar axis. For brevity, we refer to this effect as directional modulation of anisotropic electron synchrotron emission (DM-AESE). 

The above effect can be estimated geometrically by neglecting lensing and bulk-motion aberration. In Cartesian coordinates with the black hole at the origin, the wave vector toward a face-on observer is $\hat{k} = (0,0,1)$. 
For a split-monopole, the field vector takes $\vec{B} \propto \left( \delta, \cos{\t}, \sin{\t}\right)$, where $\delta$ is the toroidal-to-poloidal field ratio.
For an emission region defined by $(90^{\circ}- \Delta_{e}) \leq  \t  \leq (90^{\circ} + \Delta_{e})$, the pitch-angle range satisfies
\bea
\cos^{-1}{\left[ \sin{\Delta_{e}}\,(1 + \delta^2)^{-1/2} \right]} \leq \a_k \leq 90^{\circ} \,.
\eea 
If the emissivity peak falls within this interval, the observed image brightens. In our model, $\Delta_e \lesssim 30^{\circ} $, and $\d \sim 0.6 - 1.5$ (inferred from Fig.~\ref{fig:thetaplot}), implying that emissions from $\mB_{\upup}2, \mL_{\updown}2$ and nearly isotropic eDFs are more easily detected. Although $\mL_{\updown}2$ and $\mB_{\upup}2$ have similar angular profiles of $j_{\nu}$, their absolute amplitudes differ, making $\mB_{\upup}2$ considerably fainter.

These analyses demonstrate that the degree of anisotropy, parameterized by the beam width $\sigma$, plays a central role in shaping the observed emission through the DM-AESE. In the extreme limit of a very narrow beam (e.g.,$\sigma = 0.01$), the electron distribution becomes so sharply peaked that, for our configuration—a near-axis viewing geometry combined with MAD magnetic fields—the dominant nonthermal emission is beamed entirely out of the line of sight. The observed flux is therefore dominated by the residual thermal component, rendering such strongly beamed cases effectively indistinguishable from model $\mT$. Accordingly, we do not consider this extreme regime in detail. In Fig.\ref{fig:flux_diffevent} of Appendix.~\ref{app:imaging}, we compare the total flux and polarization fraction between model $\mT$ and a hybrid thermal+beam model with $\sigma = 0.01$, further supporting this interpretation.

\begin{figure*}[htbp]
	\centering
	\includegraphics[width=0.36\textwidth]{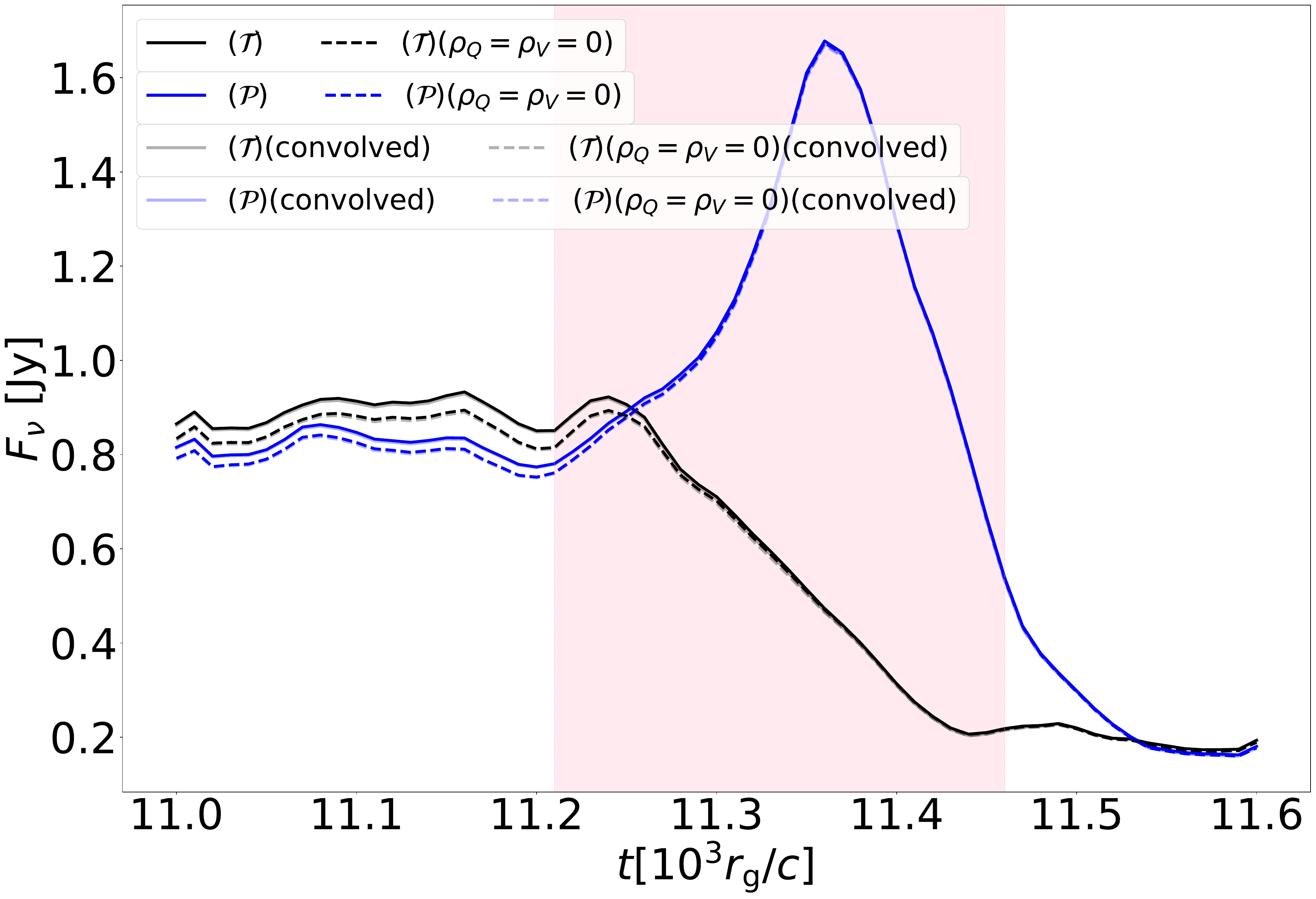} \,
	\includegraphics[width=0.36\textwidth]{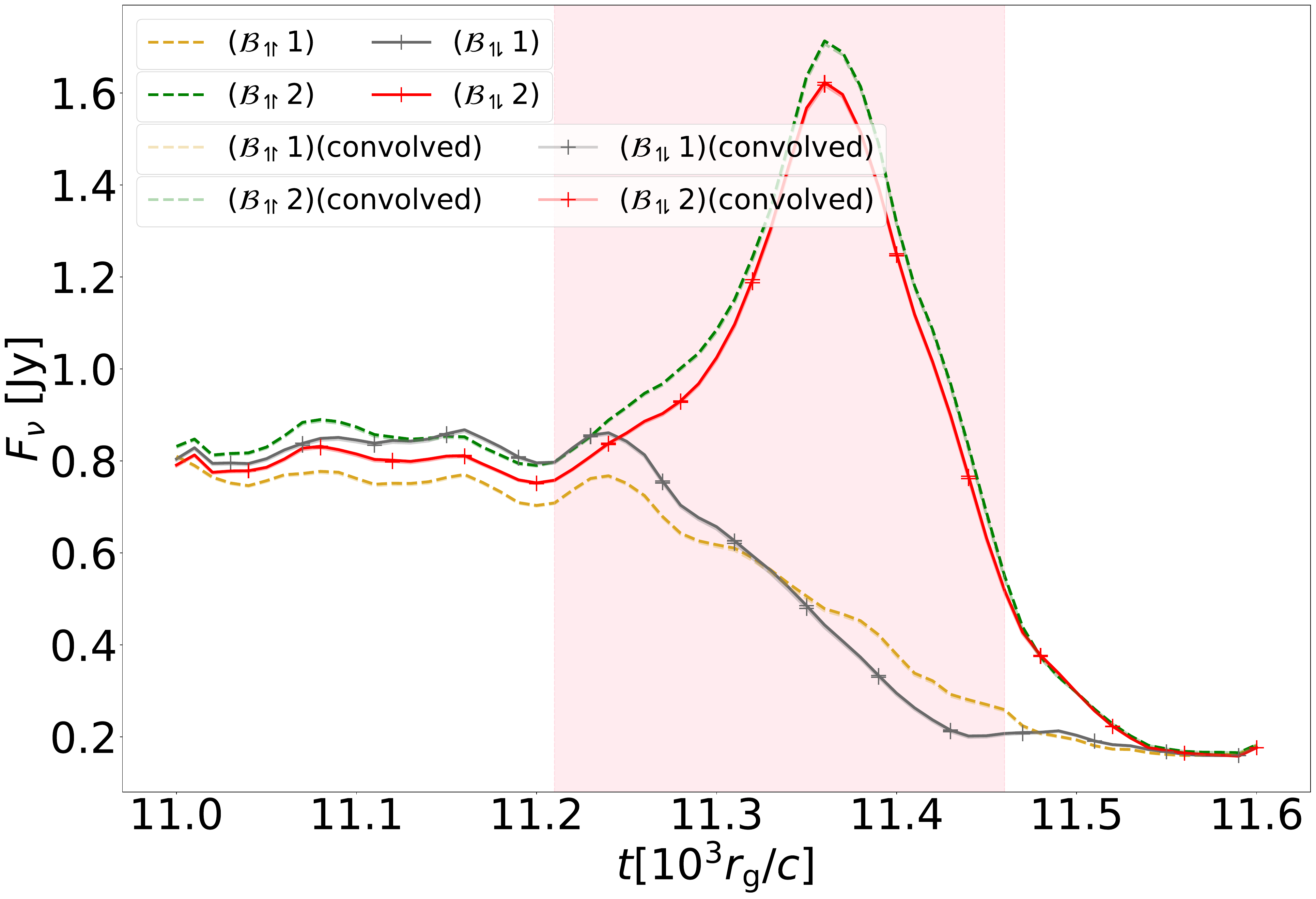} \\
	\includegraphics[width=0.36\textwidth]{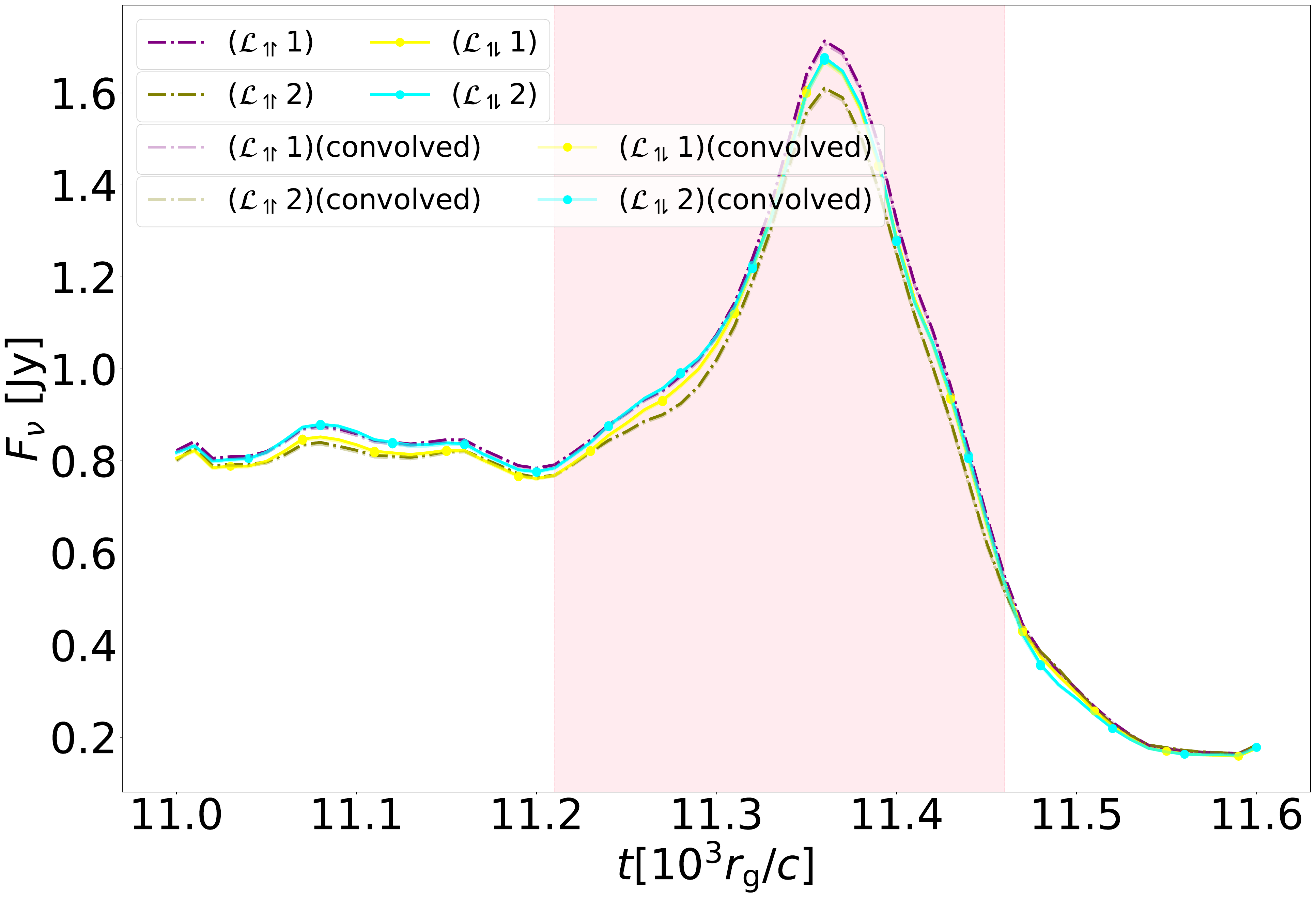} \,
	\includegraphics[width=0.36\textwidth]{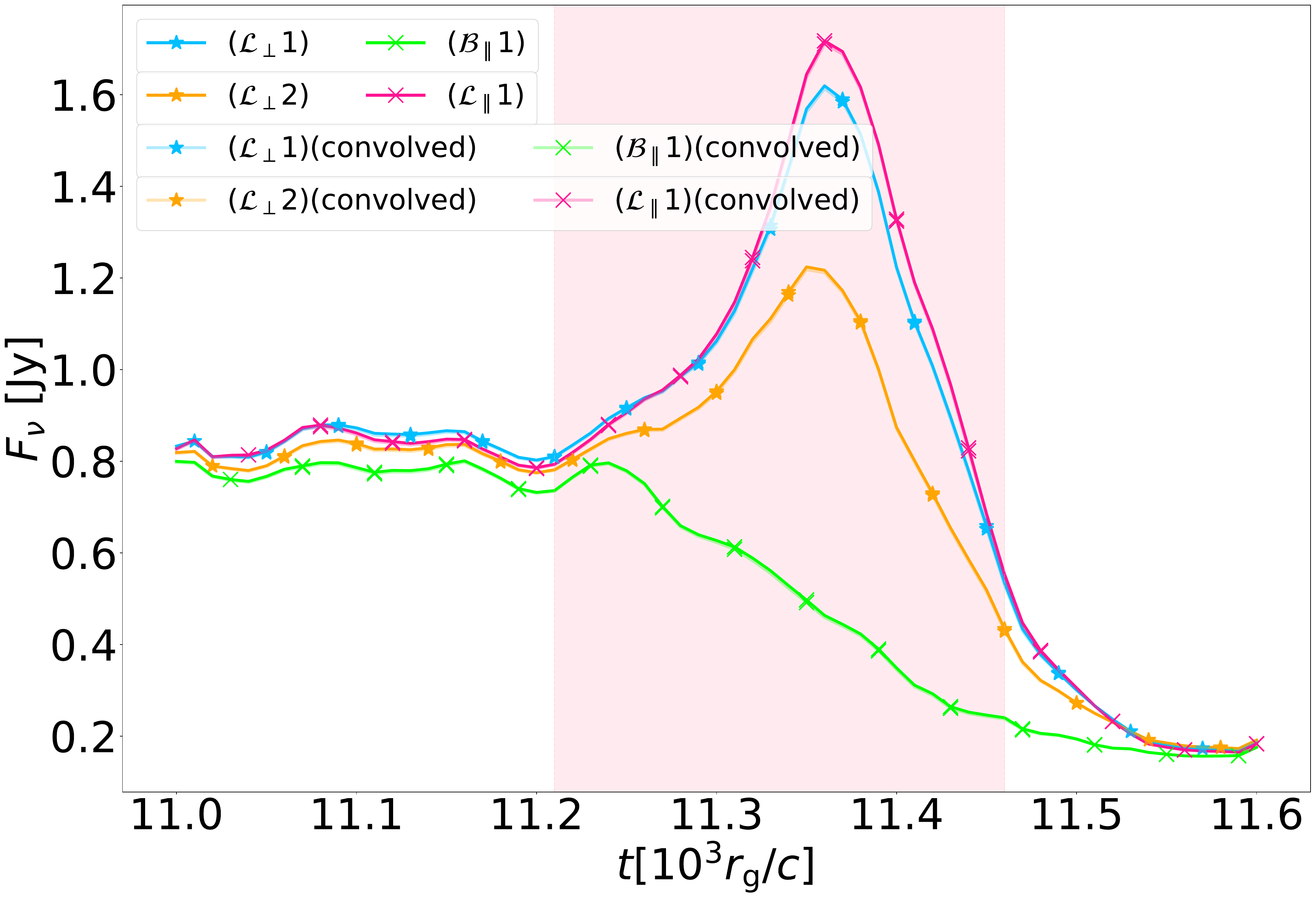}
	\caption{Time evolution of 230 GHz luminous flux for different eDF models, shown with and without convolution. Pink bands indicate the third magnetic flux eruption event.}
	\label{fig:flux}
\end{figure*}

\subsubsection{Effect of non-thermal electrons}\label{sec:effectofnonther}

Fig.~\ref{fig:flux} shows the time evolution of the  230\,GHz total flux. 
During the eruption, the flux behavior diverges significantly across models. 
If only thermal electrons are considered ($\mT$), the flux exhibits a pronounced decrease of up to approximately $75\%$, reflecting the lack of electrons in the low-density region, evacuated by outward magnetic pressure, even though electrons are heated. 
The fluxes for the strongly beamed cases $\mB_{\upup}1$, $\mB_{\updown}1$ and $\mB_{\parallel}1$ exhibit a similar behavior. In these configurations, the non-thermal emission is suppressed not only by the mismatch between the peaks of $G(\a)$ and $P^e_{\nu}(\a)$, but also by the misalignment between the dominant emission direction and the photon wave vector, i.e., DM–AESE. Consequently, highly anisotropic non-thermal electrons make only a minor contribution to the total flux.

In contrast, introducing isotropic or moderately anisotropic non-thermal electrons clearly enhances the total flux during the eruption. As shown in Fig.~\ref{fig:flux}, the models $\mP$ (and thus $\mL_{\upup}1$, $\mL_{\parallel}1$, $\mL_{\updown}1$), $\mB_{\upup}2$, $\mB_{\updown}2$, $\mL_{\upup}2$, $\mL_{\updown}2$, and $\mL_{\perp}1$ all exhibit a sharp flux rise followed by a rapid decline, peaking at nearly 1.9\,Jy at $t \approx 11360 \,t_\text{g}$, and dropping to about $0.5$\,Jy by $t \approx 11460 \,t_\text{g}$. 
Model $\mL_{\perp}$2 also produces a flare, though weaker than $\mP$, due to moderate DM–AESE, as indicated in Fig.~\ref{fig:emissivity}. 
These results suggest that during eruptions, the associated population of reconnection-driven, high-energy non-thermal electrons can indeed generate flux enhancements, potentially corresponding to an observed flare state. 

In all cases, the post-eruption flux falls below its pre-eruption level. This decline mainly results from reduced magnetization and electron temperature, combined with the disk density not yet recovered, leading to fewer non-thermal electrons, as indicated by Eq.~\eqref{ratioR}. 
Although the eDF treatment is phenomenological here, it captures the key physics of the decay phase, during which non-thermal electrons cool radiatively and collisionally, gradually returning toward a thermal state.

\subsection{Spatially resolved images}\label{sec:spatiallimage}

\begin{figure*}[htbp]
	\centering
	\textbf{(a) Intrinsic result} \\
	\includegraphics[width=0.8\textwidth]{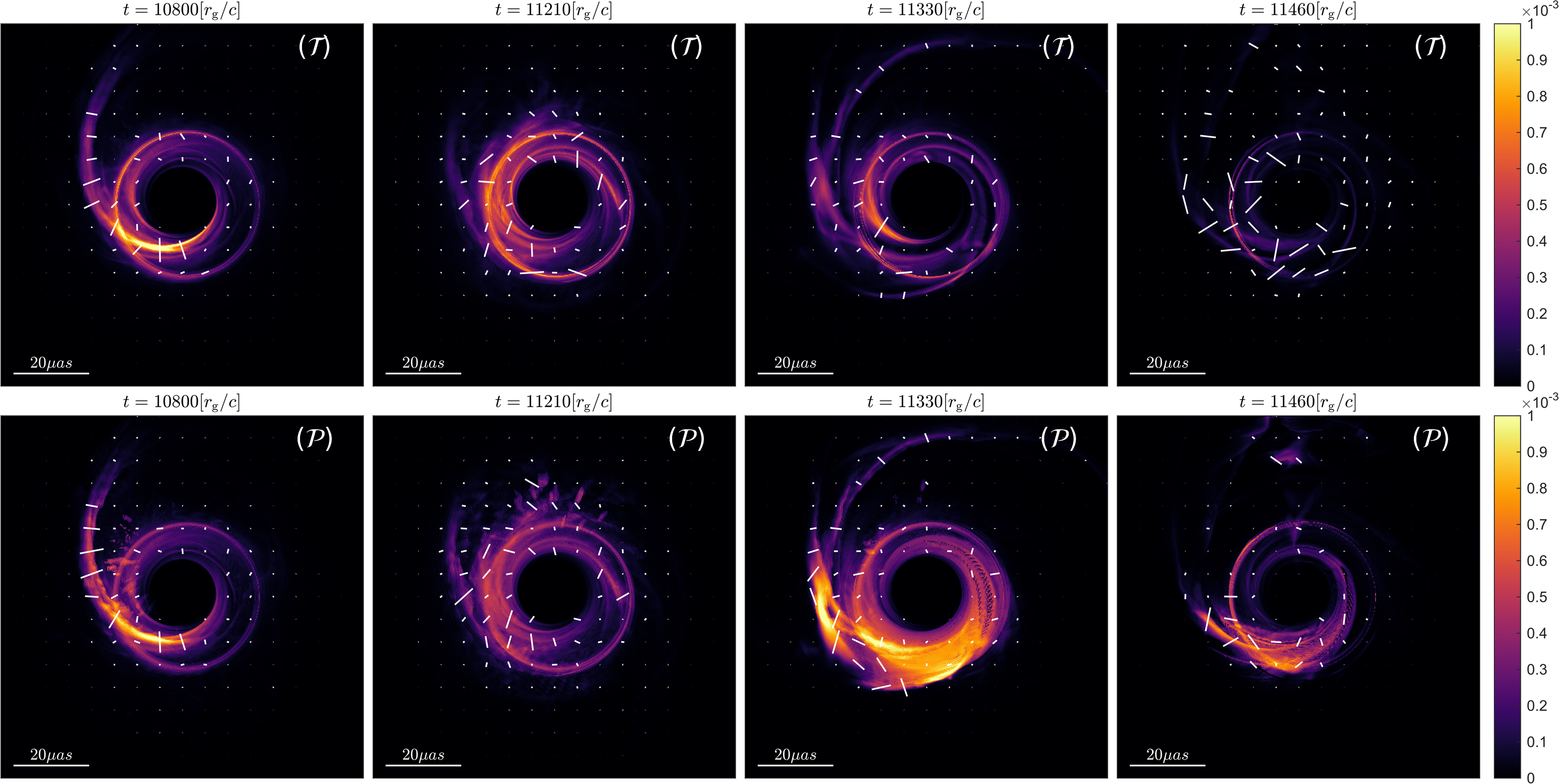} \\
	\vspace{0.5em}
	\textbf{(b) Convolved result} \\
	\includegraphics[width=0.8\textwidth]{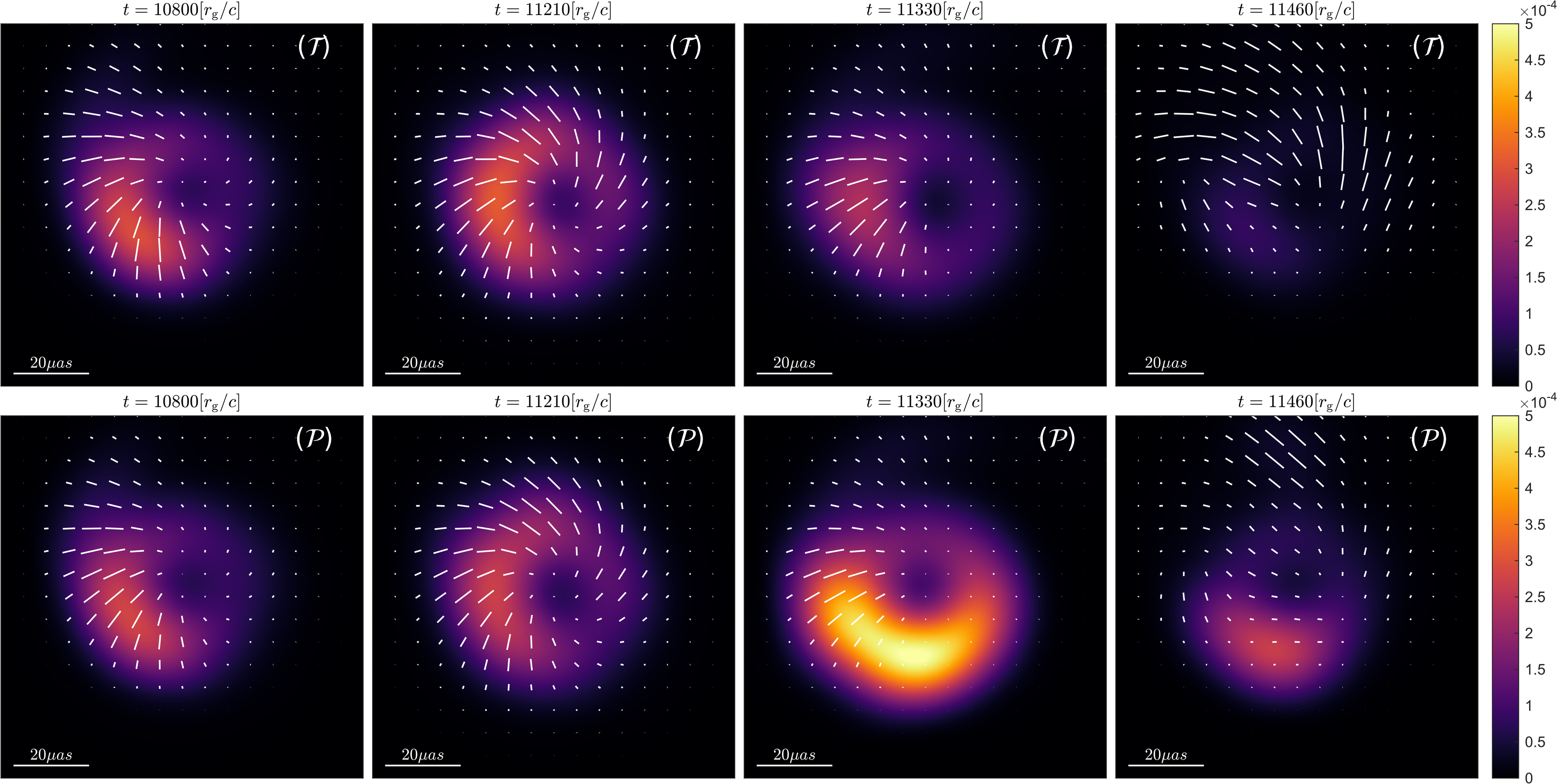}
	\caption{Intensity maps overlaid with linear polarizations at 230 GHz.
		\textbf{(a)} Results without convolution: the thermal model $\mT$ (\textbf{top row}) and the hybrid model $\mP$ (\textbf{bottom row}). 
		\textbf{(b)} Corresponding results after convolution with a Gaussian kernel (FWHM = 17 $\mu$as).
		All panels are evaluated at four time instances: $t=10800 \,t_\text{g}$, $11210 \,t_\text{g}$, $11330 \,t_\text{g}$, and $11460 \,t_\text{g}$ (columns from left to right). The unit of the intensity is $\text{erg} \,\,\text{s}^{-1} \text{cm}^{-2} \text{sr}^{-1} \text{Hz}^{-1}$.
	}
	\label{fig:thermalevpa}
\end{figure*}

\begin{figure*}[htbp]
	\centering
	\includegraphics[width=0.8\textwidth]{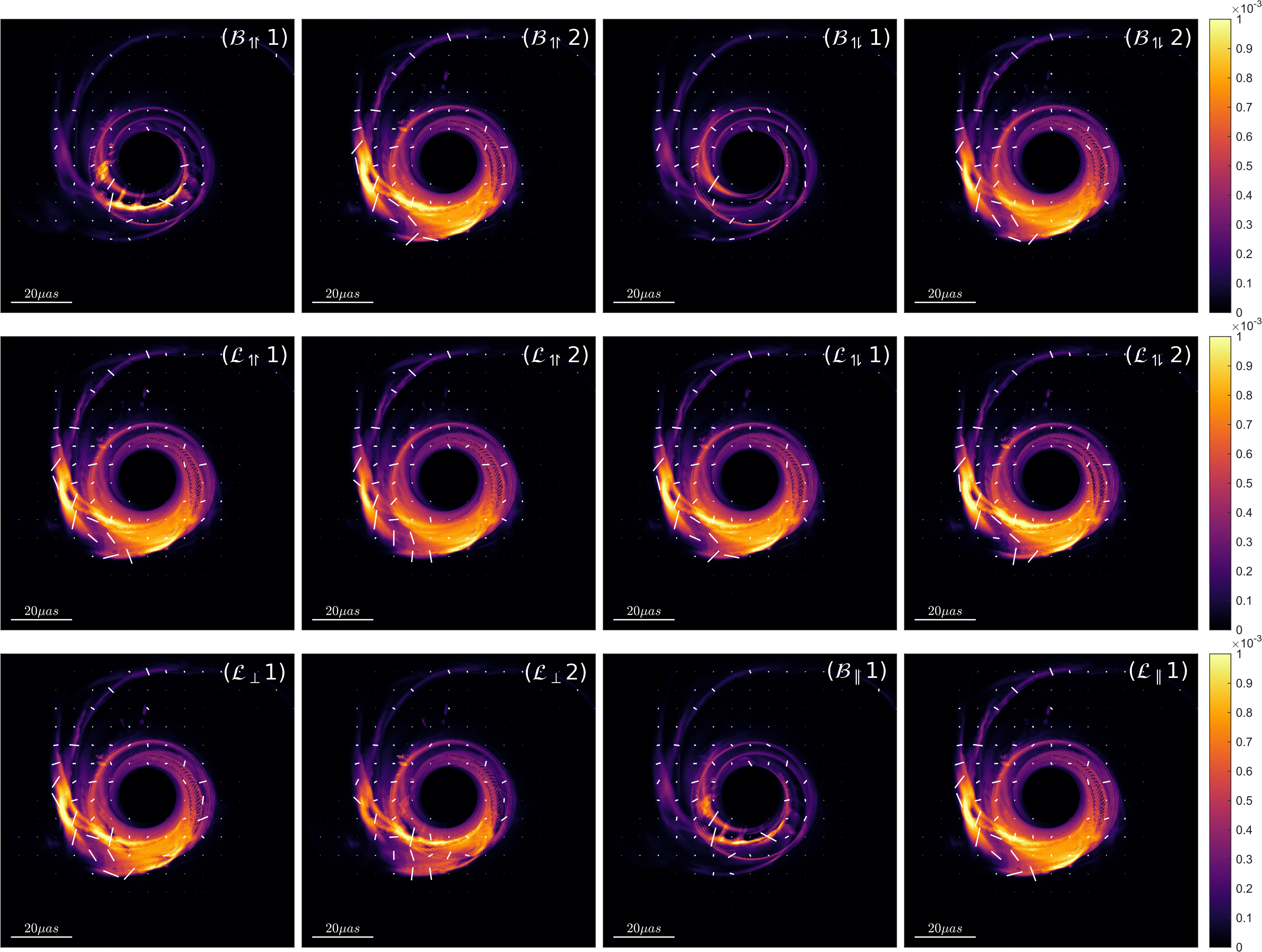}
	\caption{Intrinsic intensity maps overlaid with linear polarizations at 230 GHz from synchrotron emission of anisotropic eDF models, evaluated at the peak eruption phase $t = 11330 \,t_\text{g}$.}
	\label{fig:beamlossevpa}
\end{figure*}

Figs.~\ref{fig:thermalevpa}, \ref{fig:beamlossevpa} display the total intensity maps (color) and linear polarization pattern (white line segments) for different eDF models during four phases: before eruption, pre-eruption, peak, and post-eruption. 
In the first two phases, the images are dominated by thermal electrons and exhibit minimal variation across models. 
During the peak phase, despite a temperature increase as indicated in Fig.~\ref{fig:vr_Te}, the purely thermal model $\mT$ produces dimmer images, mainly due to the density decrease. 

When non-thermal electrons are included, the accretion flow develops a localized brightening in the lower half-plane. This feature arises from the low-density region with enhanced non-thermal emissivity, where the temperature, magnetization, and magnetic field rotation are elevated.
In the post-eruption phase, as the temperature and magnetization decrease, both $\mT$ and $\mP$ models demonstrate reduced brightness, compared to the pre-eruption phase. 
For anisotropic eDF models, the peak brightness is comparable to isotropic non-thermal electrons, except for $\mB_{\upup}1$, $\mB_{\updown}1$, and $\mB_{\parallel}1$, which exhibit lower brightness similar to the pure thermal case, consistent with their total flux behavior. 
In these field-aligned beaming scenarios, the overall suppression is linked to the predominantly toroidal magnetic geometry, which disfavors beamed emission toward the near-axis observer. At the eruption peak, although the expelled low-density region is dominated by non-vertical magnetic fields, localized and intermittent patches of enhanced vertical field emerge along its periphery. These patches temporarily align favorably with the observer's line of sight, activating the anisotropic emission and producing discontinuous bright features (blobs) superposed on a dimmed background. A direct visualization of the evolving non-thermal electron fraction and magnetic-field orientation is provided in Appendix~\ref{app:grmhd} (Fig.~\ref{fig:vr_Te}), supporting this interpretation and linking the image morphology to the underlying eruption dynamics and the flux-expulsion picture in GRMHD simulations \citep{Ripperda_2022}.

The effect of beam convolution, applied to mimic the finite resolution of the EHT, is evident from the comparison between panels (a) and (b) of Fig.\ref{fig:thermalevpa}, where the thermal ($\mT$) and hybrid ($\mP$) models are used as representative cases. While the large-scale morphology (e.g., the asymmetric brightening of the southern crescent) is largely preserved, smoothing with a 17~$\mu$as beam suppresses fine-scale structures, including substantial blurring of the photon-ring substructure and a reduction of the sharp intensity gradients present in the intrinsic images. More importantly, convolution modifies the observed linear polarization patterns: polarized streaks in the intrinsic images are averaged over the beam area, yielding smoother and more coherent polarization vectors that trace the large-scale magnetic-field geometry, as illustrated in the bottom panels of Fig.~\ref{fig:thermalevpa}.

In plots without convolution, the photon ring structure is observed, with a diameter of approximately $40\,\upmu$as. This feature is governed by extreme light bending near the photon sphere and has been extensively studied under various gravitational and astrophysical scenarios. The shape and self-similar substructure of sub-rings precisely encode the spacetime information, yet challenging to detect, requiring Earth–Moon–scale baselines \citep{Johnson:2019ljv, Lupsasca:2024xhq, Farah:2025kpb}. 
In MADs, the plasma self-absorption is non-negligible in the magnetized, high-latitude emission layer \citep{Moscibrodzka:2013rta, Davelaar_2023}. 
After an additional orbital loop, the ray's intensity is further attenuated. 
As a result, the observed emission is dominated by a brighter direct image formed during the final passage through the main emitting layer, while higher-order images produced by multiple crossings appear increasingly suppressed.

In addition, frame dragging in a rotating spacetime gives rise to distinctive observational signatures. Near the horizon, both the flow streamlines and the magnetic field become strongly twisted. In particular, the azimuthal winding increases without bound as the event horizon is approached \citep{Chen:2024jkm}. Following a local brightening triggered during an eruption episode, the emission is advected inward, forming a luminous, spiraling streamline, as shown in Fig.~\ref{fig:beamlossevpa}. Such spiral morphology encodes the imprint of spacetime frame dragging and may therefore provide a probe of near-horizon physics \citep{Ricarte:2022wpd, Hou:2024qqo, Hou:2023bep, Zhang:2024lsf}.

\subsection{Polarization patterns}

In this section, we study the linear polarization (LP) fraction and axisymmetry of the synchrotron emission during the flux eruption, especially their time variabilities.

\begin{figure*}[htbp]
	\centering
\includegraphics[width=0.8\textwidth]{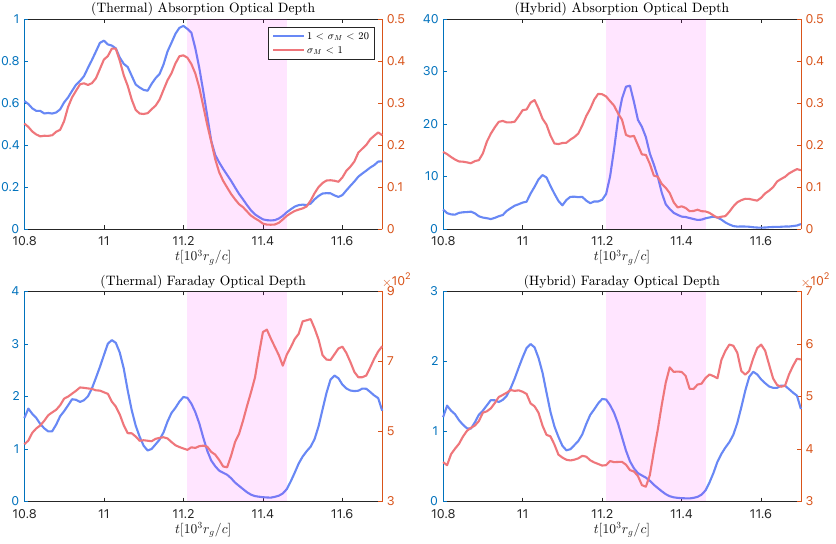}
	\caption{Time evolution of the intensity-weighted, image-plane–averaged absorption optical depth (\textbf{top}) and Faraday optical depth (\textbf{bottom}) in the thermal ($\mT$) and hybrid ($\mP$) models. For spatial comparison, the optical depths are integrated separately over segments of each ray that traverse regions with $1< \sigma_\text{M} < 20$ and with $\sigma_\text{M} < 1$. The purple bands mark the third flux-eruption event.}
	\label{depth}
\end{figure*}

\subsubsection{Optical depths}

We first outline how electron emission shapes LP evolution in an accreting plasma. The LP fraction and angle are influenced both by Faraday rotation and plasma's own emission and absorption. By defining 
the local LP fraction and angle as $m = \Mc{I}^{-1}\sqrt{\Mc{Q}^2+\Mc{U}^2}$, $\psi = \arctan{\left(\Mc{U}/\Mc{Q}\right)}$ with $\Mc{I}$ the total intensity, the transfer equation for the linear-part Stokes parameters $ \{\mathcal{Q}$, $\mathcal{U}\}$ can be written approximately as
\bea
\begin{aligned}\label{rotateeq}
	&\f{d m}{d\lambda} \approx -\f{j_I}{\Mc{I}} m + \left[\f{j_P}{\Mc{I}} - \a_P  \left( 1-m^2 \right) \right]\cos{(\psi - \psi_P)}  
	\,, \\
	&\f{d\psi}{d\lambda} \approx \rho_V - \f{1}{m} \left( \f{j_I}{\Mc{I}}-\a_P \right) \sin{(\psi - \psi_P)} \,,
\end{aligned}
\eea
where the LP emissivity and absorptivity are given by $j_P = \sqrt{j_Q^2 + j_U^2}$, $\psi_P = \arctan{\left(j_U/j_Q\right)}$, $\a_P = \sqrt{\a_Q^2 + \a_U^2}$; $j_I$ is the total emissivity, and $\rho_V$ the Faraday rotation coefficient. Circular polarization $\Mc{V}$ and the Faraday conversion between $\Mc{Q}, \Mc{U}$ and $\Mc{V}$ are weak \citep{M87_9}, and are neglected here.

For a non-emitting medium, the evolution reduces to the simple relations  $d m/ d\lambda = 0$, $d\psi/d\lambda = \rho_V$. Once emission and self-absorption are included, $\psi$ acquires an additional rotation, and the mismatch $\psi_P \neq \psi$, together with terms involving $j_I, j_P, \a_P$, generally suppresses $m$, as indicated by Eq.~\eqref{rotateeq}.
For instance, even when $\rho_V = 0$ and the solution $\psi = \psi_P$, the LP fraction evolves according to $dm/d\lambda =  -\left( m - j_P/j_I\right)j_I/\Mc{I} -\a_P \left( 1-m^2 \right) $, where absorption always decreases $m$, while emission drives it toward $j_P/j_I$.
This emission (absorption)-induced reduction of $m$ provides an important depolarization channel along the ray, which we refer to as dichroic depolarization for short \citep{1979Lightman}.

To assess the optical depths of the rays forming our synthetic images, we plot the image-plane–averaged absorption depth, $\tau_A = \int \a_I d\lambda$, and Faraday depth, $\tau_F = \int |\rho_V| d\lambda$ in Fig.~\ref{depth}.
The average is weighted by the total intensity at each image-plane pixel. For spatial comparison, we  integrate the optical depths separately along segments of each ray that pass through regions with $1< \sigma_\text{M} < 20$ and with $\sigma_\text{M} < 1$.
The results show that synchrotron emission and absorption arise primarily in magnetized regions with $\sigma_\text{M} > 1$, where the Faraday depth is negligible. In contrast, the weakly magnetized, near-equatorial zone ($1< \sigma_\text{M} < 20$) emits little but produces extremely strong Faraday rotation. Together, these features reveal an accretion flow composed of a cold, dense midplane that supplies strong Faraday screening bracketed by hotter, high-latitude emission layers \citep{Moscibrodzka:2017gdx, Ricarte:2020llx, Davelaar_2023}. 
The emission layers in the thermal model $\mT$ remain optically thin throughout the evolution, whereas the hybrid thermal–non-thermal model $\mP$ develops higher absorption depth and becomes optically thick during the eruption episode, indicating non-negligible depolarization from dichroic depolarization.

\subsubsection{Linear polarization fraction}

There are two variables to quantify the averaged LP fraction on the image plane, namely the unresolved, image-integrated LP fractions $m_{\text{net}}$, as well as their resolved, image-averaged counterparts $\langle |m| \rangle$:
\bea
\begin{aligned}\label{eq:averm}
&m_{\text{net}} = \frac{\sqrt{\left(\sum_i \Mc{Q}_i\right)^2+\left(\sum_i \Mc{U}_i\right)^2}}{\sum_i \Mc{I}_i} \,, \\
&\langle |m| \rangle = \frac{\sum_i \sqrt{\Mc{Q}_i^2+\Mc{U}_i^2}}{\sum_i \Mc{I}_i} \,,
\end{aligned}
\eea
where $i$ denotes a pixel on the image plane. 
Note that $m_{\text{net}}$ is insensitive to observational resolution, whereas $\langle |m| \rangle$ is not \citep{M87_7, M87_8}. 
In synthetic imaging studies, $\langle |m| \rangle$ will converge as the pixel number increases \citep{Pihajoki:2018ihj} and thus can well reflect the resolved LP degree on the observer's image plane.

\begin{figure*}[htbp]
	\centering
	\includegraphics[width=0.38\textwidth]{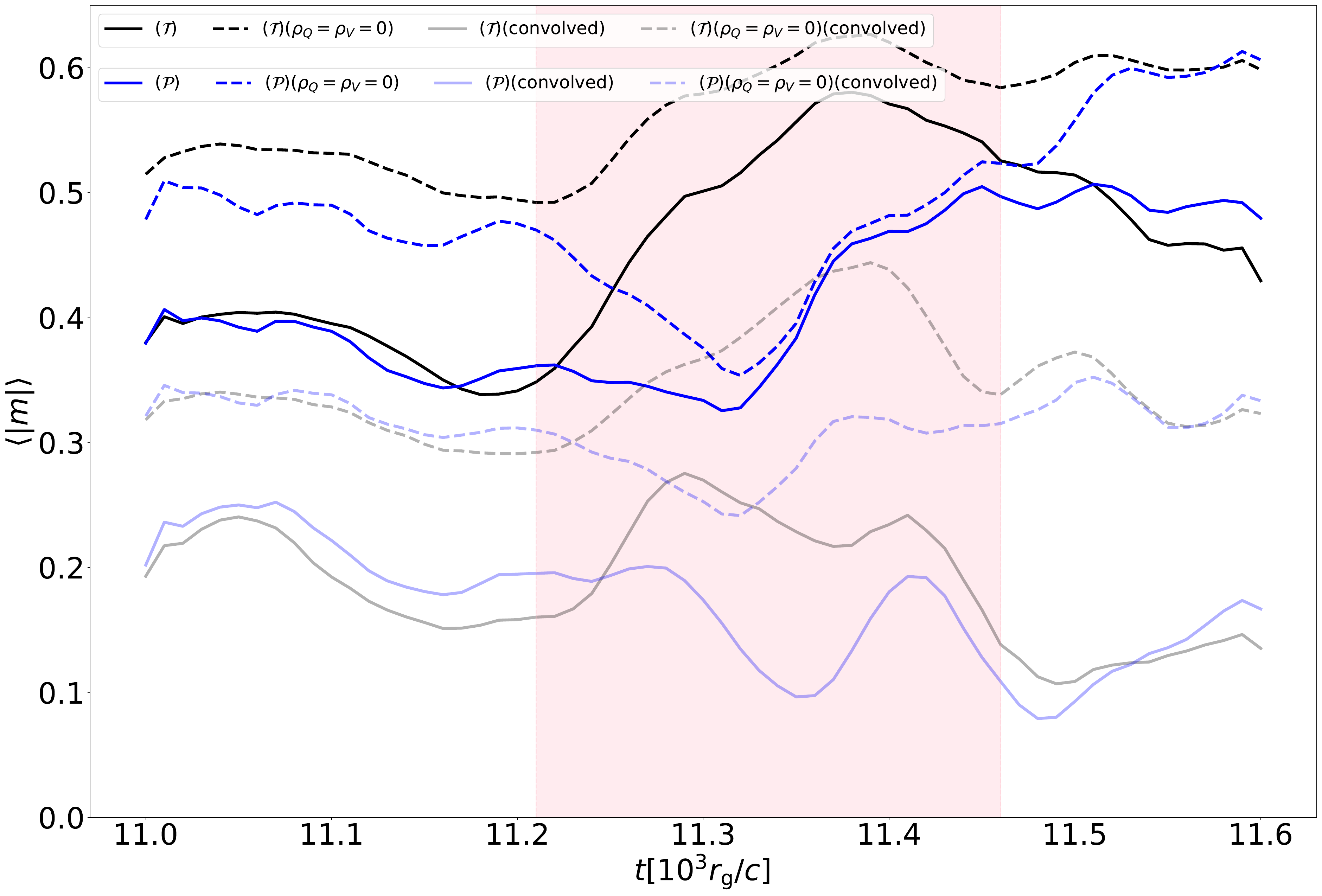}\,
	\includegraphics[width=0.38\textwidth]{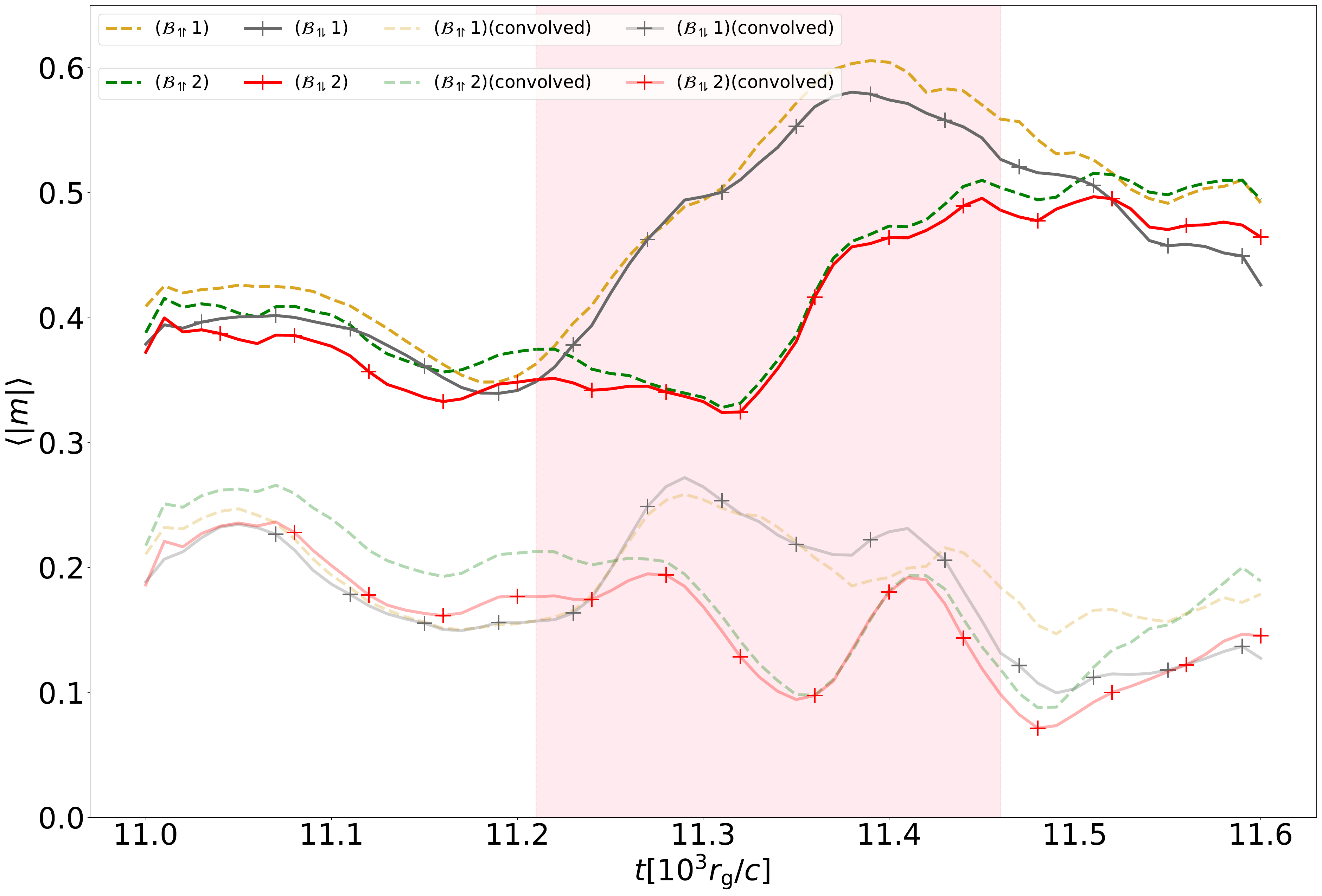}\\
	\includegraphics[width=0.38\textwidth]{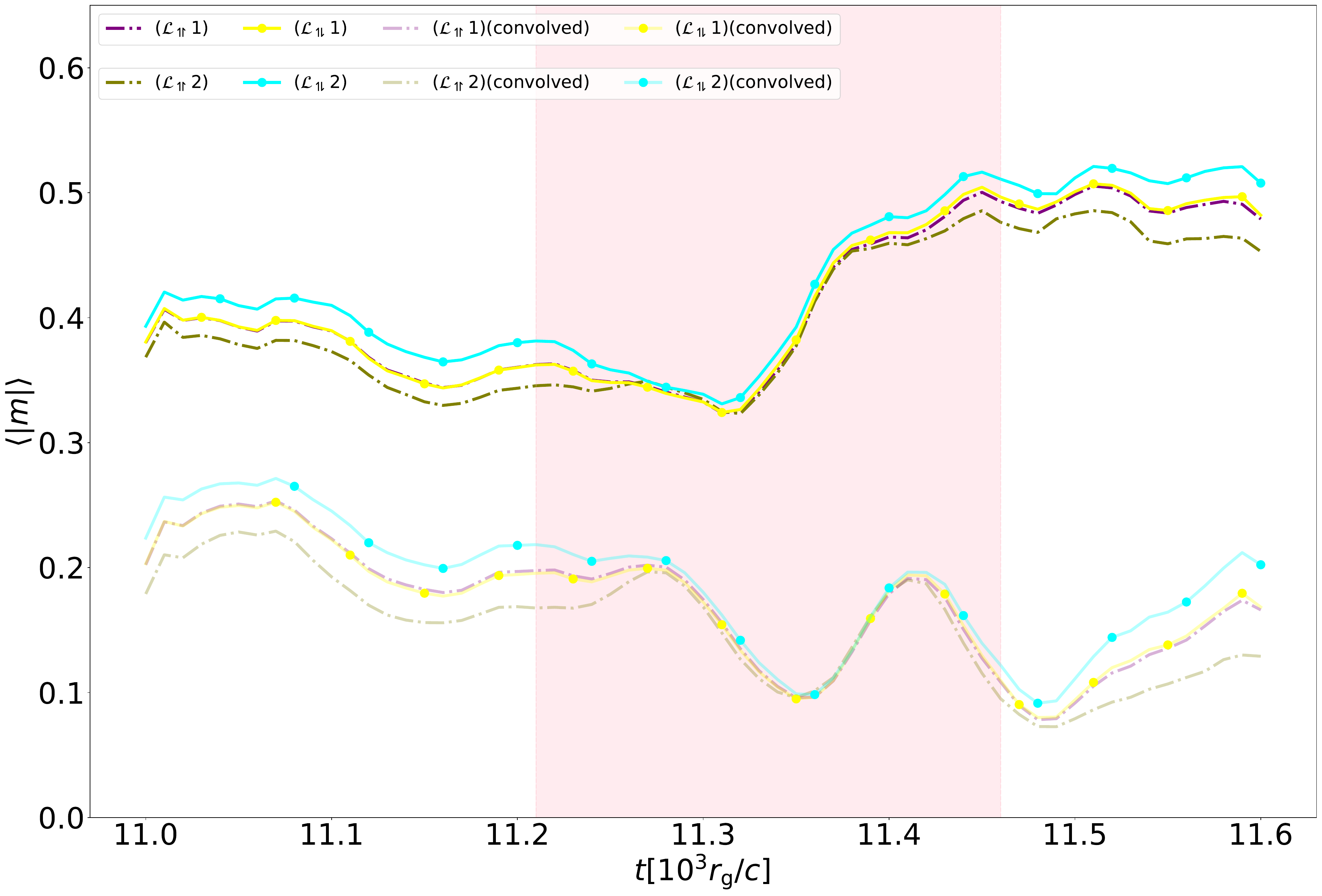}\,
	\includegraphics[width=0.38\textwidth]{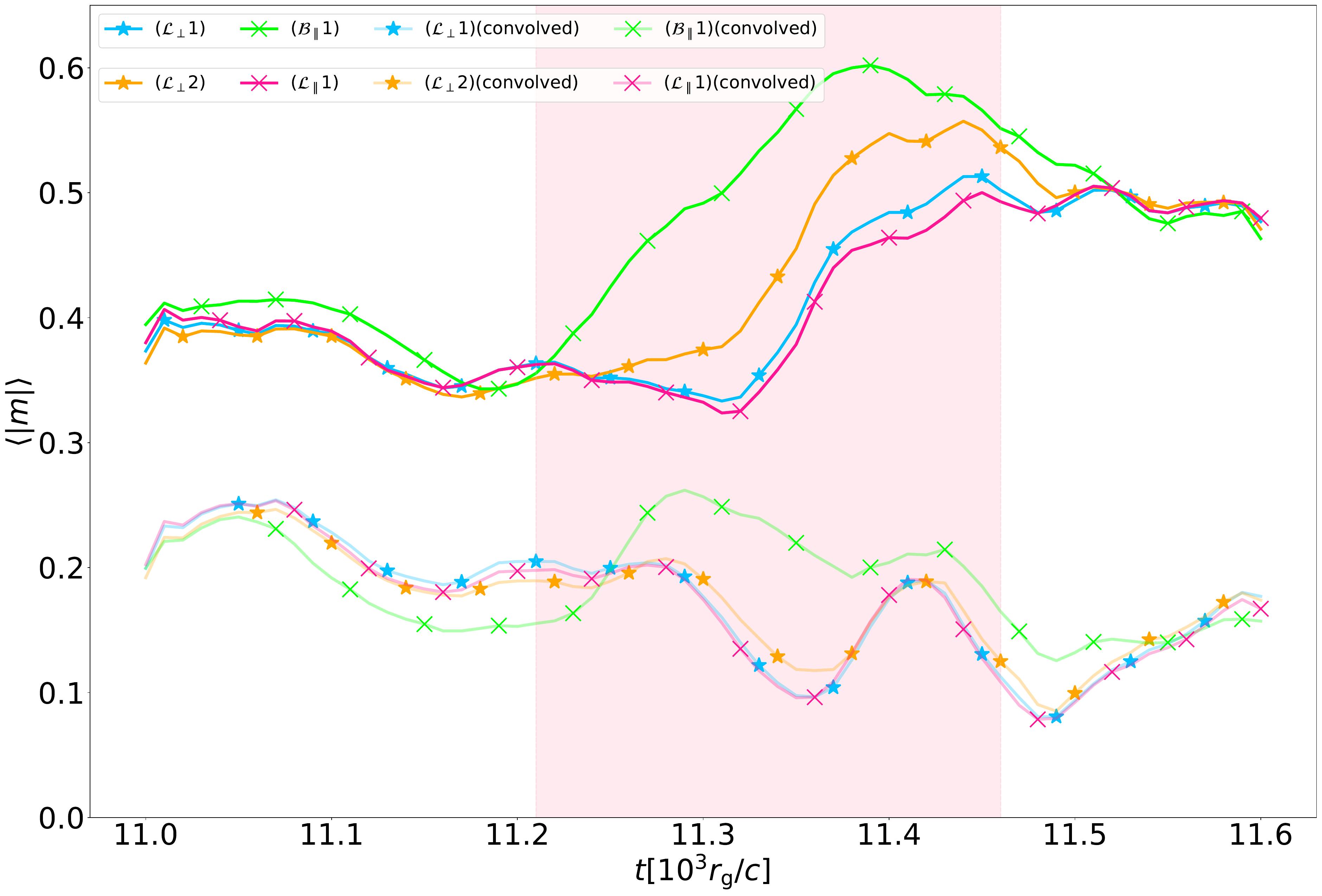}
	\caption{Time evolution of the LP fraction $\langle |m| \rangle$, evaluated for different eDF models, shown with and without convolution. Pink bands indicate the third flux eruption event.}\label{fig:polarization}
\end{figure*}

We plot $\langle |m| \rangle$ for different eDF models in Fig.~\ref{fig:polarization}, showing both the intrinsic (unconvolved) results and those convolved with a circular Gaussian kernel of FWHM $=17\,\upmu$as. We first focus on the intrinsic results.
Before the eruption, when the fraction of non-thermal electrons is small, the models show minor differences, and $\langle |m| \rangle$ remains at $35\% \sim 40\%$.  
During the eruption, we observe an enhancement in $\langle |m| \rangle$ for all models.  
In particular, for the $\mT$, $\mB_{\parallel}1$, $\mB_{\upup}1$ and $\mB_{\updown}1$ models, $\langle |m| \rangle$ reaches about $60\%$ near $t = 11400 \,t_\text{g}$. 
The increase can be attributed to the expulsion of matter by the outward magnetic pressure, which reduces emissivity and self-absorption (Fig.~\ref{depth}), thus reducing the dichroic depolarization along a light ray.

For the model $\mP$, $\langle |m| \rangle$ exhibits a plateau before a subsequent enhancement, with the transition occurring near $t = 11360 \,t_\text{g}$---coinciding with the peak of the total flux. 
It is partially attributed to the transition between optically thin and thick states \citep{Tsunetoe:2024uzh}, as indicated in Fig.~\ref{depth}.
During the earlier stage of the eruption, the absorption becomes strong as a result of the enhancement of non-thermal electrons emission. Thus, the plasma becomes optically thick with a typical local LP degree of $\Pi = 3(6p+13)^{-1}$, which is smaller than that of optically thin case,  where $\Pi = (p + 1)(p+7/3)^{-1}$ \citep{1979Lightman} \footnote{Besides, according to Eq.~\eqref{eq:PICp}, an increase in magnetization leads to a smaller power-law index, which also results in a decrease/increase in $\Pi$ for optically thin/thick plasma. For anisotropic eDFs, the intrinsic LP fraction should be modified by the energy dependence of the electron anisotropy \citep{Comisso_2023}. }. In the later stage of eruption, the plasma returns to optically thin and the LP degree increases clearly. 

The top-left panel also compares the cases with zero and nonzero Faraday coefficients. We find that the Faraday effect suppresses $\langle |m| \rangle$, particularly prior to the eruption. From Eq.~\eqref{rotateeq}, one infers that a larger $\rho_V$ increases the mismatch between $\psi$ and $\psi_P$, thereby strengthening the dichroic depolarization.
Conversely, when $\rho_V$ is very small, $\psi$ and $\psi_P$ remain nearly aligned throughout propagation, which greatly limits depolarization. This behavior is also reflected in the enhanced LP fraction at the end of the eruption, where the Faraday optical depth approaches zero.

For most of anisotropic-electron models, their polarization behavior still tends to resemble that of the $\mT$ or $\mP$ cases, analogous to the situation for the flux. 
However, both the flux and polarization fraction of $\mL_{\perp}2$ lie between those of $\mT$ and $\mP$, further indicating that its corresponding electron anisotropy produces the most prominent observational signatures.

Due to beam depolarization, i.e., averaging over small-scale EVPA variations, the convolved pre-eruption LP fractions in all models decrease to $15\% \sim 25\%$, closer to the observed values. During the eruption, models with substantial non-thermal contributions (e.g., $\mP$) maintain systematically and significantly lower convolved LP fractions than the purely thermal model ($\mT$). As discussed for the intrinsic results, enhanced non-thermal emission increases the local optical depth and Faraday rotation near the eruption peak, producing a more disordered polarization pattern. When smoothed with a Gaussian beam, this leads to stronger cancellation of polarized substructures. After the eruption, the convolved LP fractions return to values comparable to their pre-eruption levels.

\subsection{Second azimuthal Fourier mode}

We further characterize the polarization structure using the secondary azimuthal Fourier mode of the linear polarization \citep{Palumbo:2020flt}. It is defined by integrating the complex Stokes parameter  $\Mc{P} = \Mc{Q} + i \Mc{U}$, weighted by an azimuthal phase factor, over an annulus on the image plane:
\begin{equation}
	\beta_2 = \frac{1}{\Mc{I}_{\rm ann}} \int_{\rho_\text{min}}^{\rho_\text{max}} \int_0^{2\pi} \Mc{P} e^{-2i\phi} \rho \, \dif \rho \, \dif \phi  \,,
\end{equation}
where $(\rho, \phi)$ are polar coordinates, $\rho_\text{min},\rho_\text{max}$ define the annulus boundaries, and $\Mc{I}_{\rm ann}$ is the total intensity within it. The magnitude $0\leq|\beta_2|\leq 1$ measures the azimuthal order of the polarization field, while the phase $\arg(\beta_2)$ encodes the dominant electric-vector position angle (EVPA) orientation \citep{Palumbo:2020flt, M87_8}.
For a narrow, axisymmetric ring viewed face-on, one recovers the quadrupolar relation $\arg(\beta_2) = 2\,\text{EVPA}\big|_{\phi = 0^{\circ}}$. 

\subsubsection{Image-integrated results}\label{sec:Fullbeta2}

\begin{figure*}[htbp]
	\centering
	\includegraphics[width=0.38\textwidth]{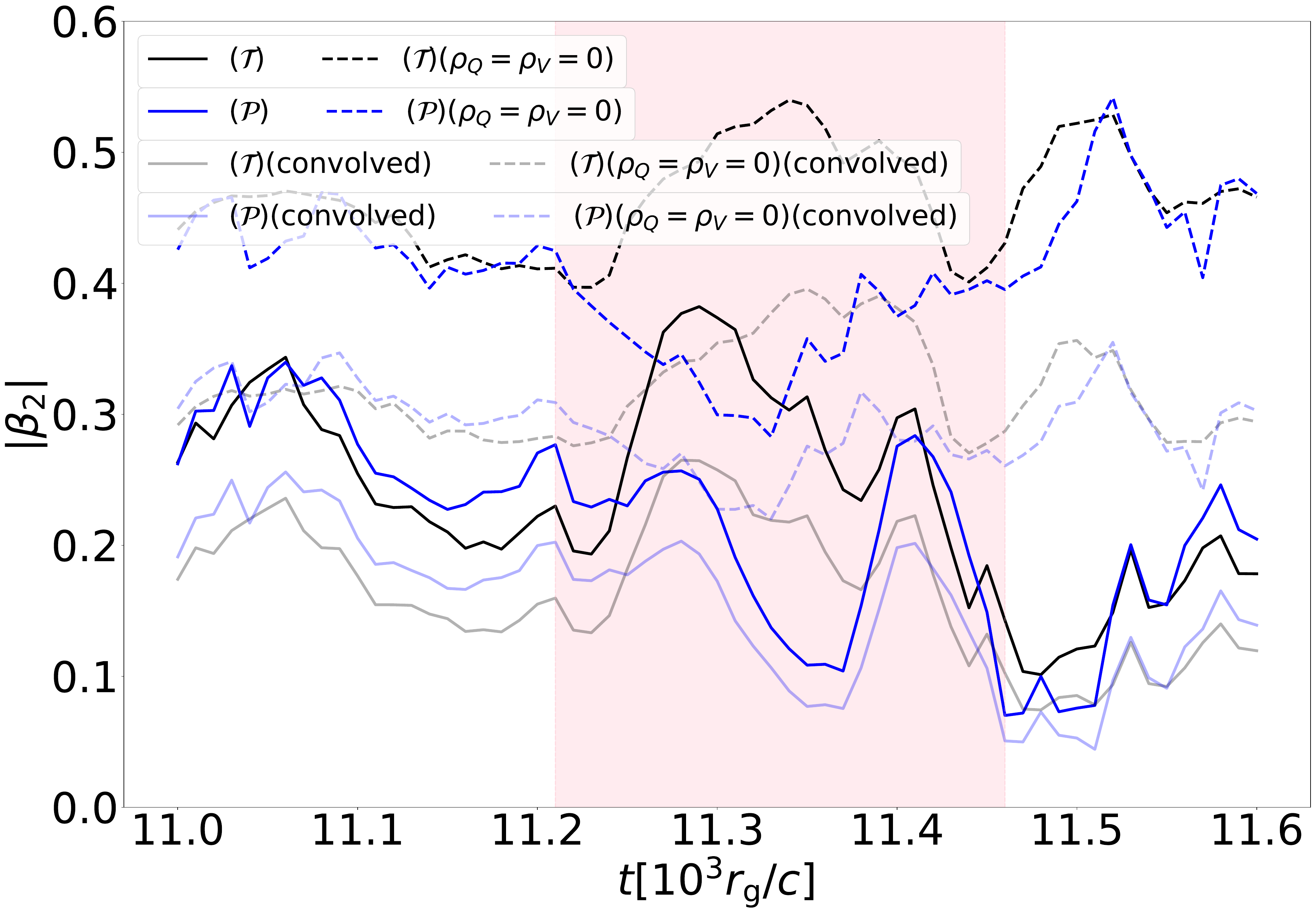}\,
	\includegraphics[width=0.38\textwidth]{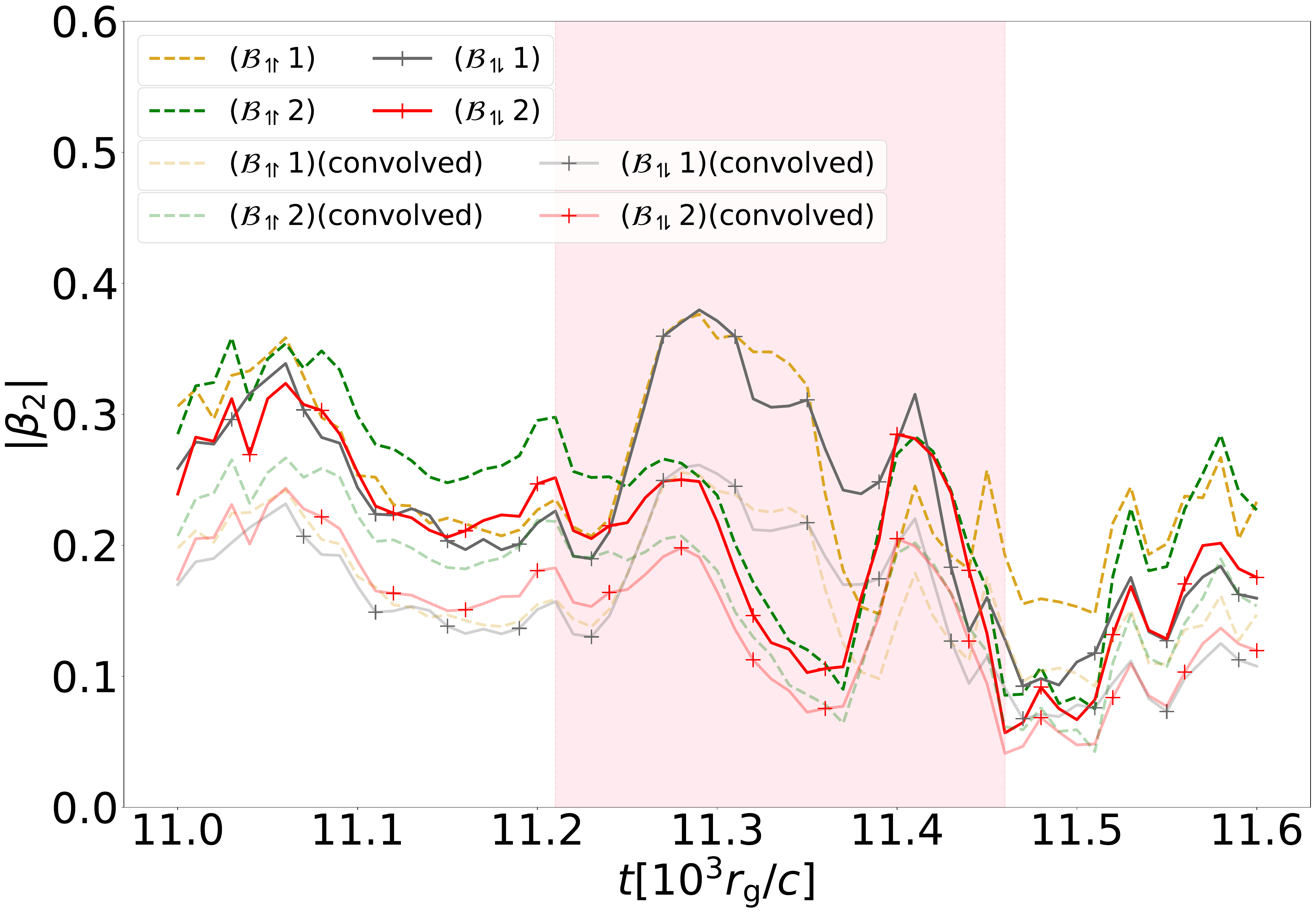}\\
	\includegraphics[width=0.38\textwidth]{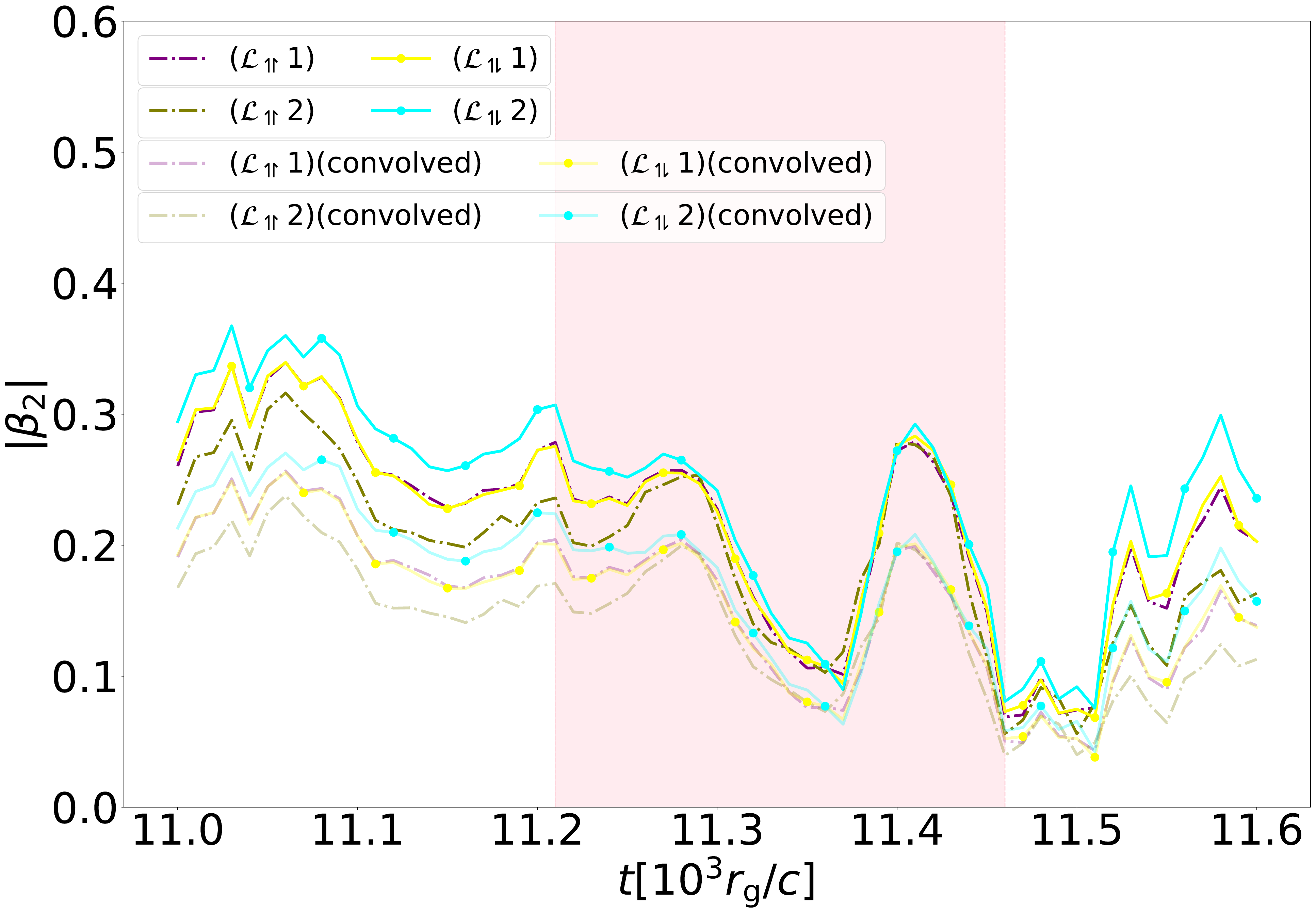}\,
	\includegraphics[width=0.38\textwidth]{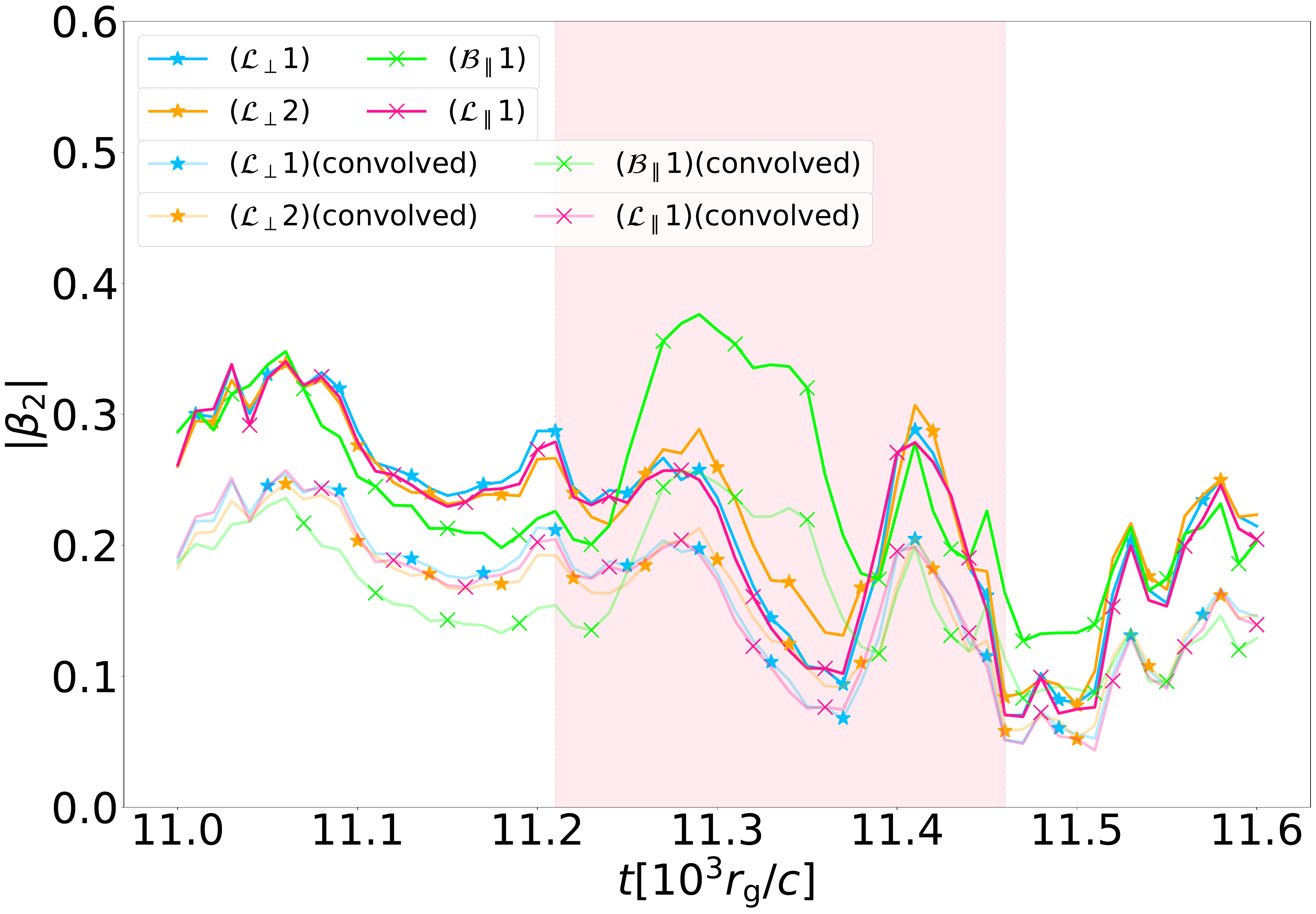}
	\caption{Time evolution of image-integrated $|\beta_2|$, evaluated for different eDF models, shown with and without convolution. Pink bands indicate the third flux eruption event.}\label{fig:beta2}
\end{figure*}

\begin{figure*}[htbp]
	\centering
	\includegraphics[width=0.38\textwidth]{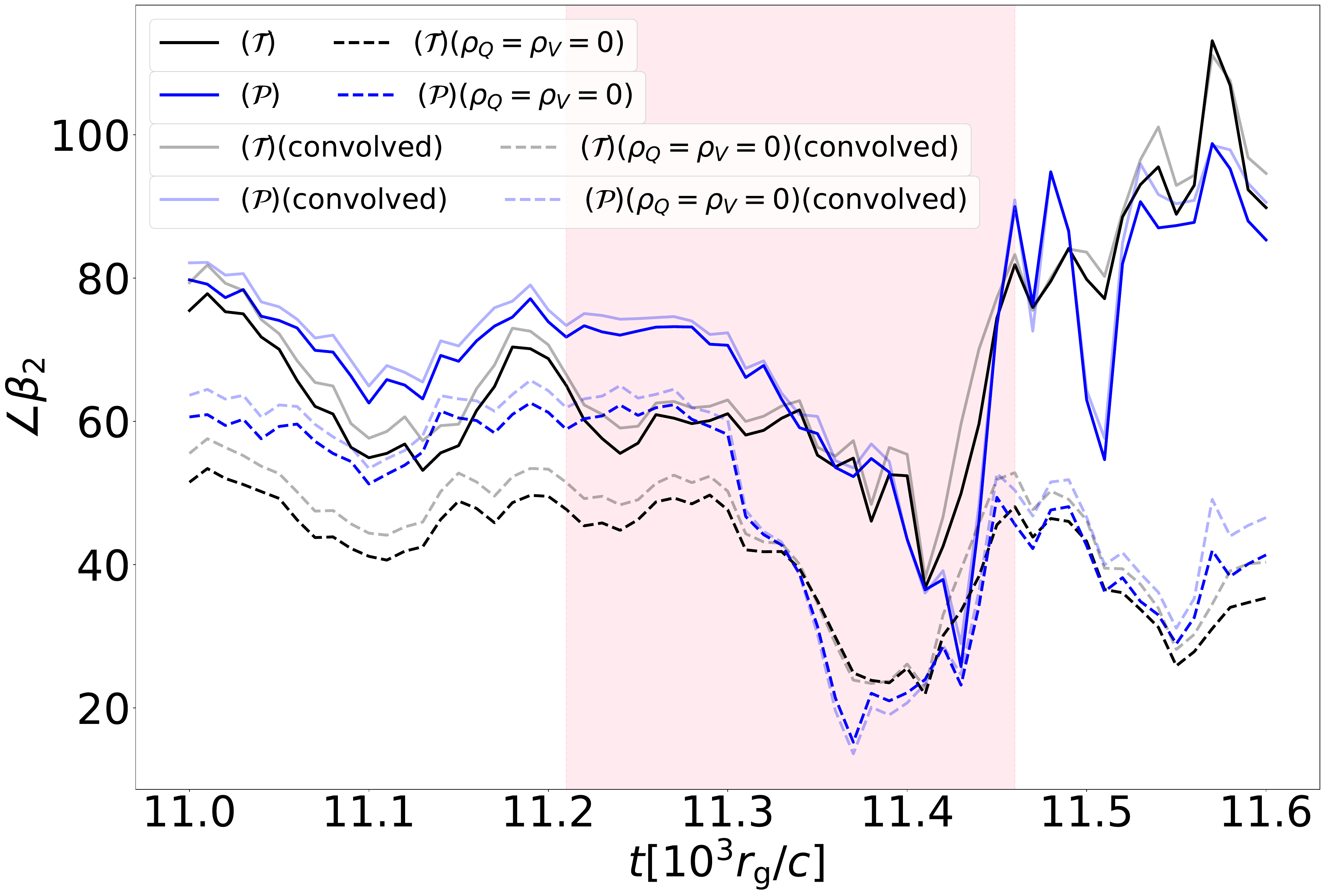}\,
	\includegraphics[width=0.38\textwidth]{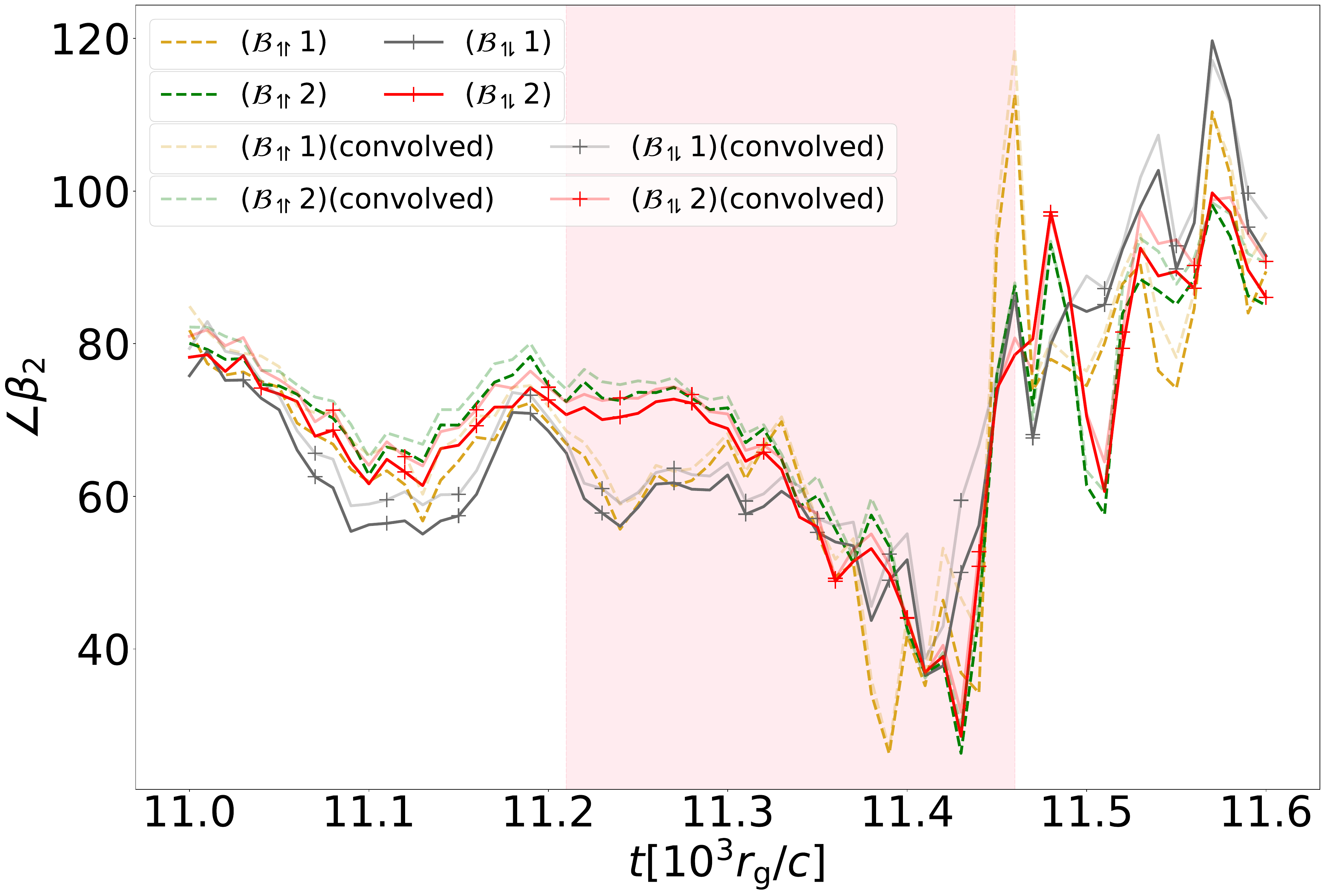}\\
	\includegraphics[width=0.38\textwidth]{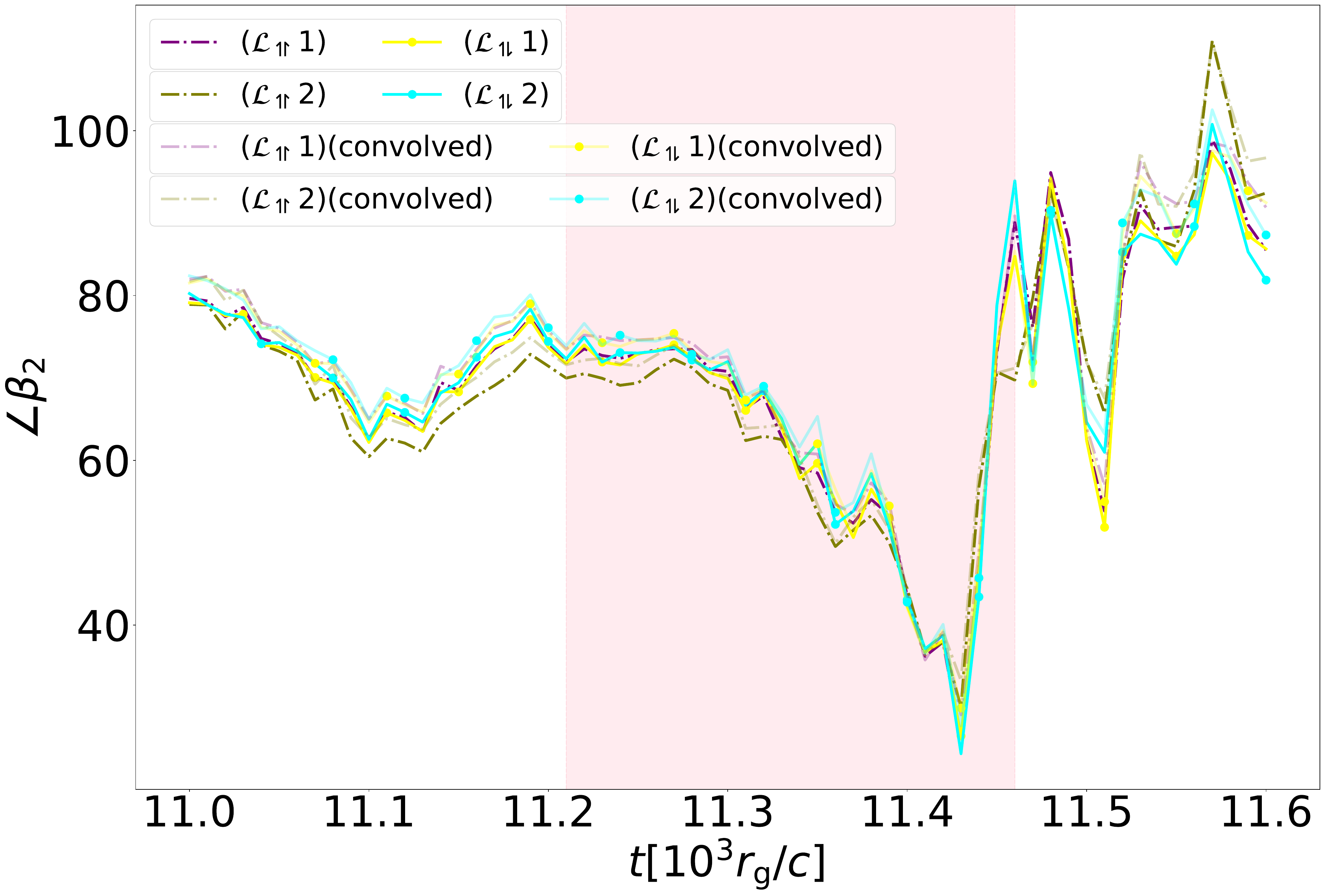}\,
	\includegraphics[width=0.38\textwidth]{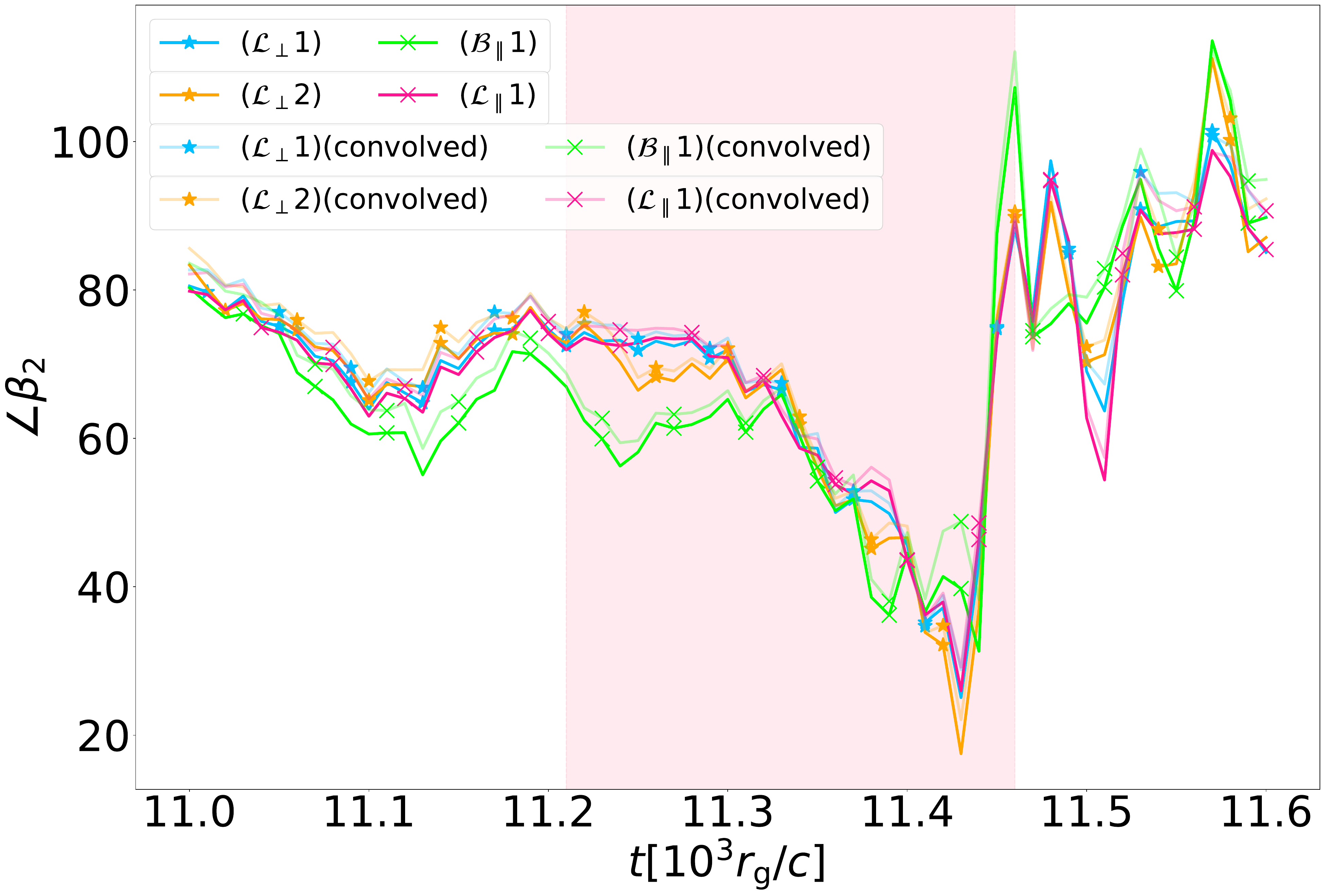}
	\caption{Time evolution of image-integrated $\arg(\beta_2)$, evaluated for different eDF models, shown with and without convolution. Pink bands indicate the third flux eruption event.}\label{fig:beta2angle}
\end{figure*}

In Figs.~\ref{fig:beta2} and \ref{fig:beta2angle}, we show the temporal evolution of $|\beta_2|$ and $\arg(\beta_2)$ integrated over the full image domain, with annulus boundaries $\rho_\text{min}=0$ and $\rho_\text{max}\to\infty$. Both unconvolved and convolved results are shown; we first focus on the former. Several distinct trends emerge. In the thermal model $\mT$, $|\beta_2|$ exhibits a pronounced peak: it rises from $0.2$ to $0.4$ at $t \approx 11280 \,t_\text{g}$, and then declines to about $0.1$ at later times. The case with zero Faraday rotation also increases during this interval, though more moderately, reflecting reduced dichroic depolarization and thus more intrinsically ordered polarization.

In the $\mP$ model, $|\beta_2|$ with and without Faraday effects both show an initial decline followed by a recovery during the eruption. 
Both minima coincide with the epoch of peak total flux. Because $|\beta_2|$ is weighted by the local intensity, its evolution is dominated by the brightening in the lower half of the image. At the flux maximum, this strongly emitting region—rich in non-thermal electrons and optically thick—produces more disordered polarized emission, leading to the observed minimum in $|\beta_2|$. 
By the end of the eruption, although the local LP fractions remain high, turbulence and deformation in the re‑accreting material increase line‑of‑sight variations in Faraday rotation, reducing the coherence of the polarization pattern and suppressing $|\beta_2|$ across all eDF models.

The phase $\arg(\beta_2)$ remains positive in all models ($20^{\circ} \sim 120^{\circ}$), consistent with previous results for prograde MAD inflows (with sign reversal expected for observers near the southern pole). 
As shown in Fig.~\ref{depth}, the dominant emission arises from the magnetized layer above the equatorial plane, where the Faraday depth is modest, so $\arg(\beta_2)$ primarily reflects the intrinsic magnetic geometry. The dense equatorial layer, though Faraday-thick, contributes negligible emission and therefore does not influence the integrated phase.

For most of the period following the onset of the eruption, $\arg(\beta_2)$ steadily declines from about $70^{\circ}$ to $30^{\circ}$ across all models, reflecting the emergence of toroidal magnetic fields in the emission region, as indicated by Fig.~\ref{fig:thetaplot}. This behavior can be understood as follows: for a near-axis observer, the projected linear polarization vectors are approximately orthogonal to the local magnetic field (neglecting plasma motion and propagation effects). A smaller EVPA therefore corresponds to a more radial polarization pattern, implying a field configuration dominated by toroidal components.

Toward the end of the eruption, $\arg(\beta_2)$ increases markedly, implying a recovery of poloidal field as well as the Faraday rotation, as can be inferred from the results of $\mT,\mP$ with and without $\rho_Q,\rho_V$.
In most eDF models, this enhancement approaches $90^{\circ}$, whereas in models that incorporate a substantial population of beam electrons streaming along field lines ( $\mB_{\parallel}1, \mB_{\upup}1$ ),  $\arg(\beta_2)$ can rise to $110^{\circ}$.

After convolution, $|\beta_2|$ follows a similar temporal evolution but is systematically reduced by $\sim 0.1$, as beam depolarization smooths small-scale polarized structures and diminishes the azimuthal coherence of $\Mc{P}$ across the image. In contrast, $\arg(\beta_2)$ remains nearly unchanged, since the phase is primarily set by the large-scale orientation of the polarization pattern, which is largely preserved under convolution.

\subsubsection{Annular decomposition}\label{sec:Annularbeta2}

\begin{figure}[htbp]
	\centering
\hspace*{-5mm}
	\includegraphics[width=0.53\textwidth]{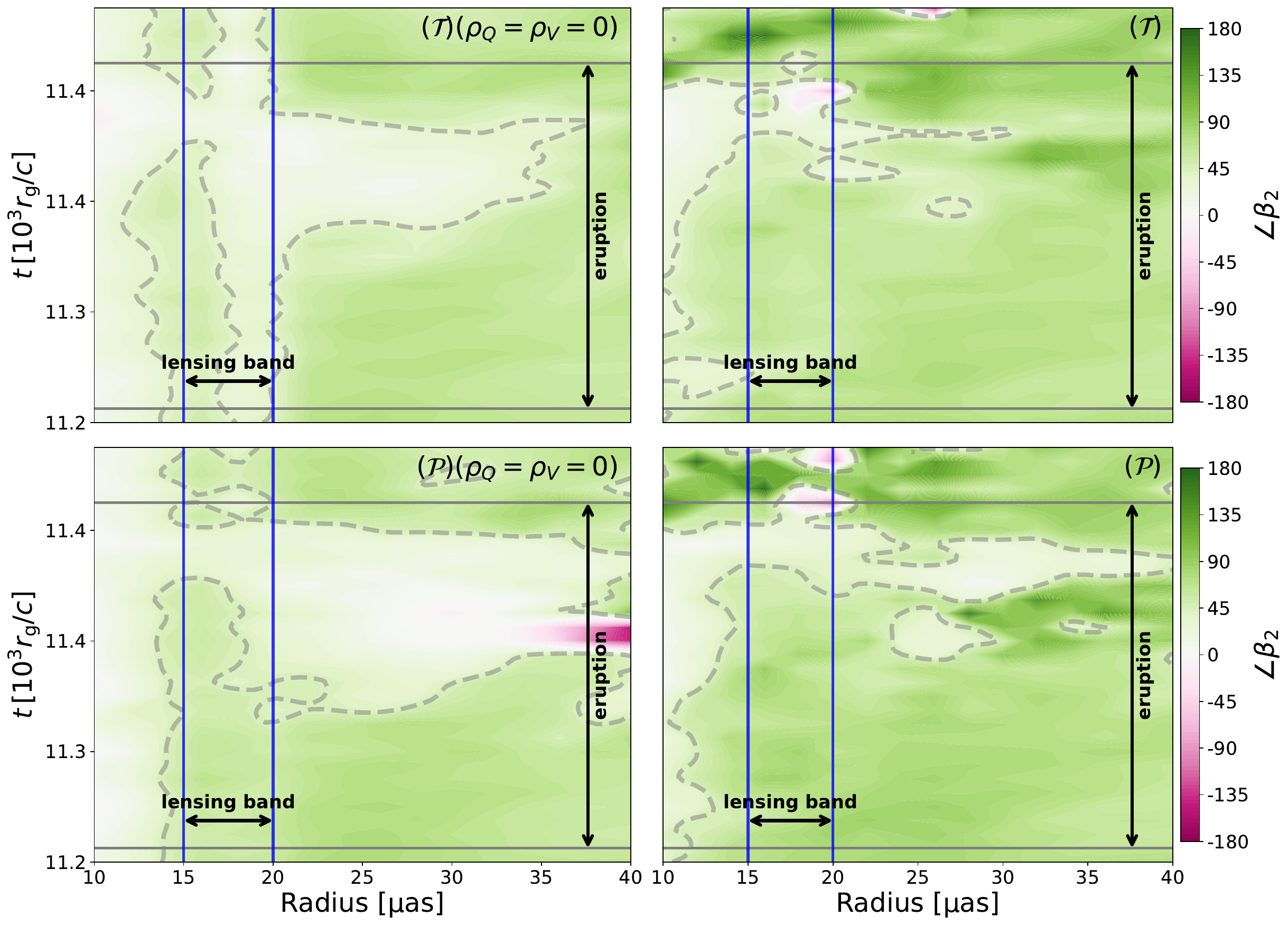}
	\caption{Intrinsic radial profiles of $\arg(\beta_2)$ for $\mT,\mP$ models, evaluated as a function of image-plane radius and examined over different evolution times. The left columns present the results with the Faraday coefficients artificially set to zero, while the right columns show the full results including Faraday effects. The blue lines delineate the lensing band, defined by null geodesics that cross the equatorial plane twice before reaching the observer. The dashed contours enclose regions where $\arg(\beta_2) \leq 40^{\circ}$ in the $\mT$ model and $\arg(\beta_2) \leq 45^{\circ}$ in the $\mP$ model.}
	\label{fig:2Dbeta2TP}
\end{figure}

Typically, the integration over $\beta_2$ is performed across the entire image plane. Here, we relax this constraint and allow $\rho_\text{min},\rho_\text{max}$ to be chosen freely, which enables us to examine the detailed polarization patterns influenced by the flux eruption.
Fig.~\ref{fig:2Dbeta2TP} presents the temporal evolution of the radial profile of $\arg(\beta_2)$ in the $\mT$ and $\mP$ models. 
Results of other eDF models do not exhibit remarkable or enlightening features, and are included in the Appendix.~\ref{app:imaging} for completeness.

When Faraday coefficients are turned off, the polarization angle exhibits two time-independent regions of steady decline in the $\mT$ model: one located at the lensing band ($16 \sim 20 \, \upmu$as),  and the other close to the event horizon ($\lesssim 15 \,\upmu$as). 

The lensing-band region corresponds to light rays that cross the equatorial plane twice, so the net polarization reflects the superposed Stokes parameters accumulated over two passages through the emitting region, corresponding to the direct and lensed images \citep{Johnson:2019ljv}. In the $\mT$ model, where the optical depth is small, both the direct and lensed images contribute appreciably, yielding a clear feature in $\arg(\beta_2)$ across the lensing band. In contrast, the much larger optical depth in the $\mP$ model suppresses the lensed-image contribution; the total polarization is dominated by the primary image, and the variation of $\arg(\beta_2)$ across the lensing band is therefore less pronounced.

When the Faraday coefficients are included, the strong equatorial Faraday screen induces substantial polarization-angle rotation along the ray path \citep{Moscibrodzka:2017gdx, Ricarte:2020llx}, and the coefficients further enhance depolarization through Eq.~\ref{rotateeq} as the ray traverses the entire plasma. As a result, the lensing‑band feature in the $\arg(\beta_2)$ profile becomes invisible for all eDF models.

In the region close to the event horizon, frame dragging generates a strong toroidal magnetic field, producing a very small EVPA for near-axis observers—a generic property of magnetized accretion flows. Because this field-dominated polarization signature originates primarily from the main emission layers through the primary image, it is only weakly sensitive to optical-depth effects, yielding similar behavior in both the $\mT$ and $\mP$ models.

\section{Summary and Discussion}
\label{sec:conclusion}

In this work, we have investigated the 230 GHz imaging and polarization signatures produced by synchrotron emission from non-thermal electrons in a MAD during a flux–eruption event. We performed a 3D GRMHD simulation of a MAD around a Kerr black hole, identifying the flux eruption by the rapid drop of the horizon-threading magnetic flux from a local maximum to a subsequent minimum (Sec.~\ref{sec:evolution}).

To analyze the emission properties, we constructed a phenomenological hybrid thermal–non-thermal eDF model (Sec.~\ref{sec:eDF}). Nonthermal electrons were assumed to be energized from the thermal pool via magnetic reconnection, forming a power-law tail whose index and energy fraction followed fitting functions from existing PIC. We have also introduced a Gaussian function with adjustable center and width to model possible pitch-angle anisotropies, allowing for beam like or loss-cone distributions (Sec.~\ref{sec:anisotropiceDF}).

We have examined the local emissivity for different eDFs and found that the combination of single-electron emission anisotropy and eDF anisotropy significantly modified the emissivity, even producing a double-cone structure (Sec.~\ref{sec:emissivityplot}). 
We then performed GRRT calculations to generate 230 GHz images for several fiducial eDF models, focusing on the evolution of total intensity and polarization throughout the flux-eruption episode (Sec.~\ref{sec:imagingresults}). The GRMHD outputs were rescaled such that the pre-eruption total flux matched 0.66 Jy for the near-axis observer with $\t_o = 17^{\circ}$, consistent with M87* observations. Our main findings are summarized as follows:

$\bullet$ Introducing a non-thermal eDF ($\mP$, $\mB_{\upup}$2, $\mB_{\updown}$2, $\mL_{\upup}$1, $\mL_{\updown}$1, $\mL_{\parallel}1$, $\mL_{\upup}$2, $\mL_{\updown}$2, $\mL_{\perp}$1, $\mL_{\perp}$2) can indeed enhance the total flux during the eruption, demonstrating that non-thermal electrons play a non-negligible role in black-hole flaring states (Sec.~\ref{sec:effectofnonther}). Even modest changes in the non-thermal fraction efficiently modify the 230 GHz emissivity.

$\bullet$ A localized brightening emerges on the image plane as the flux outburst develops, corresponding to low-density, high-temperature, highly magnetized regions with higher non-thermal electron population (Sec.~\ref{sec:spatiallimage}). This brightening also leaves imprints on the image-integrated pattern of $|\beta_2|$ (Sec.~\ref{sec:Fullbeta2}).

$\bullet$ Moderate eDF anisotropy has only a minor impact on the emission profile, leaving both the flux and polarized image nearly indistinguishable from those of the isotropic nonthermal model $\mP$. In contrast, strongly anisotropic eDFs (such as $\mB_{\upup}1$, $\mB_{\updown}1$, $\mB_{\parallel}1$) can misalign the peak emission direction with the line of sight, thereby suppressing nonthermal emission toward the near-axis observer; as a result, the 230 GHz image morphology becomes largely degenerate with that of the thermal model $\mT$ (Sec.~\ref{sec:effectofaniso}). 

$\bullet$ The linear-polarization fraction in the thermal model ($\mT$) increases during eruption due to reduced density and optical depth, whereas that in the hybrid model ($\mP$) decreases, as enhanced non-thermal emissivity raises the local optical depth, producing strong dichroic depolarization (Sec.~\ref{sec:evolution}). 

$\bullet$ The image-integrated polarization orientation $\arg(\beta_2)$ decreases during eruptions in all models, reflecting the growing dominance of the toroidal field component (Sec.~\ref{sec:evolution}). 
In the radial profile of $\arg(\beta_2)$, the strong-lensing signature is blurred by optical depth and by the equatorial Faraday screen, especially in model $\mP$ model, where the emission region is nearly optically thick. 
The frame-dragging–induced near-horizon feature remains potentially observable despite in-medium effects (Sec.~\ref{sec:Annularbeta2}). 

$\bullet$ After convolution with a circular Gaussian beam of FWHM $=17\,\upmu$as, enabling direct comparison with current EHT observations, the pre-eruption LP fractions in all models decrease to $\sim 15\%$--$25\%$, and the variability during the eruption is smoothed. Correspondingly, $|\beta_2|$ is reduced by $\sim 0.1$ while $\arg(\beta_2)$ remains nearly unchanged. This suggests that $\arg(\beta_2)$ may serve as a more robust diagnostic of the accretion state than other observables. Joint constraints on the accretion flow from multiple polarization observables will require higher angular resolution.

In conclusion, our results show that both the non-thermal electron population and its anisotropy play distinct roles in shaping the observed variability during MAD flux-eruption events. 
Non-thermal electrons can drive flux outbursts and localized brightening on the image plane, while reducing the linear-polarization fraction by increasing the absorptive optical depth along the line of sight. 
The anisotropy of the eDF reshapes the angular distribution of the intrinsic emissivity and modulates the observability of non-thermal electrons. 
Together, these effects provide a coherent and physically motivated framework for diagnosing the accretion state and probing the underlying plasma processes through light-curve and polarimetric observations \citep{EventHorizonTelescope:2024uoo}.

Several physical processes lie beyond the scope of this study. Under the fast-light approximation, emission is projected instantaneously onto the image plane, neglecting light-travel–time differences. Including these delays in a slow-light treatment introduces a characteristic lag of order $\sim \pi\tilde{r}_0/c$ between the lensed and direct images, with $\tilde{r}_0\sim 3M$ the photon-sphere radius related to face-on observers. Rapid fluid variability may also induce radiation perturbations that imprint observable substructure in the direct image.
Multi-frequency predictions (e.g., extending to 86 GHz) and their dependence on viewing angle are essential for connecting with future high-sensitivity, long-baseline observations (e.g., the ngEHT). Observations at 86 GHz are particularly valuable because, unlike at 230 GHz where the thermal core can remain competitive, the intrinsic synchrotron emission at lower frequencies is dominated by the non-thermal power-law tail across both optically thin and thick regimes. Consequently, 86,GHz data are far more sensitive to the spatial distribution and dynamics of non-thermal electrons.
Moreover, while anisotropic distributions can appear partially degenerate with isotropic or purely thermal models for near-axis observers (e.g., M87*), larger inclination angles can align the line of sight more favorably with beamed non-thermal emission. This helps break the degeneracy, allowing the distinctive signatures of pitch-angle anisotropy to emerge more clearly. A dedicated survey over frequency and inclination is deferred to future work.

Although anisotropy plays a minor role in the flux and polarization for near-axis observers such as at $\t_o = 17^{\circ}$, this does not imply that its influence remains small at larger viewing angles, where the observer could intercept more directional non-thermal emission. For example, a beam-like distribution peaks in local emissivity at angles of $\sim 30^{\circ}$ relative to the magnetic field, which can imprint distinctive signatures on the image plane at appropriate inclinations and may even permit inference of the underlying magnetic-field geometry.

\section*{Acknowledgments}

We thank Yosuke Mizuno and Hongxuan Jiang for their assistance and insightful discussions. We also thank Ye Shen for his initial investigations. The work is partly supported by NSFC Grant No. 12275004, 12205013, 12575048, 12547123 and 12547127. M. Guo is also supported by Open Fund of Key Laboratory of Multiscale Spin Physics (Ministry of Education), Beijing Normal University.

\appendix

\section{Supplementary Material for the GRMHD Simulation}\label{app:grmhd}

\begin{figure*}[htbp]
	\centering
	\includegraphics[width=0.85\textwidth]{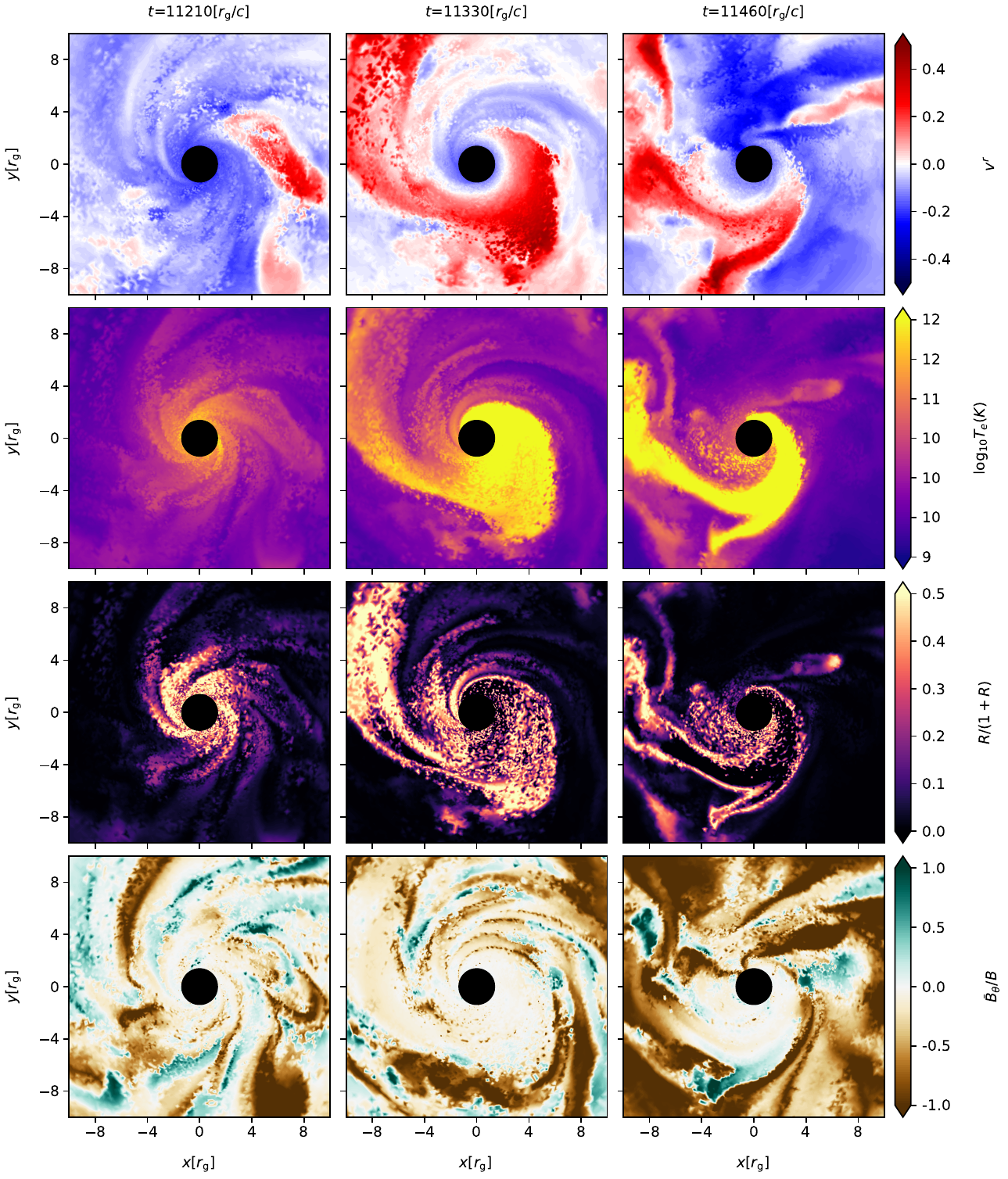}
	\caption{Distribution of radial velocity $v^r = u^r / u^t$ (\textbf{the first row}), electron temperature $T_\text{e}$ (\textbf{the second row}), non-thermal electron fraction $R/(1+R)$ (\textbf{the third row}) and fractional vertical magnetic-field component $\bar{B}_\theta/B$ (the fourth row) in the $x-y$ plane at $t = 11210 \,t_\text{g}$, $t = 11330 \,t_\text{g}$ and $t = 11460 \,t_\text{g}$.}
	\label{fig:vr_Te}
\end{figure*}

In this section, we present supplementary figures from the GRMHD simulations that can support the discussion in the main text.
The first row of Fig.~\ref{fig:vr_Te} shows the radial velocity in the $x-y$ plane. Before the eruption, plasma at the inner disk edge accretes at roughly $0.2 c$, slower than free fall --- a hallmark of the MAD state \citep{Narayan2003}. 
During the eruption, magnetic-energy release imparts substantial kinetic energy to the plasma, which is expelled outward at about $0.3 c$, driving turbulence in the low-density region.
The second row of Fig.~\ref{fig:vr_Te} demonstrates distribution of electron temperature $T_e$ in the $x-y$ plane. At $t = 11210\, t_\text{g}$, electron temperatures in the inner disk reach $T_\text{e} \sim 10^{11}$K, produced by the conversion of gravitational potential energy into kinetic and internal energy during accretion. At the eruption peak ($t = 11330\, t_\text{g}$), significant heat fluxes ($T_\text{e} \sim 10^{12}$K) appear along magnetic field lines (black arrows), especially within the low-density region. These mainly arise from the conversion of magnetic energy into kinetic and thermal energy, consistent with the concurrent decrease in magnetic flux in Fig.~\ref{fig:mdot}. By $t = 11460 \, t_\text{g}$, these heat fluxes dissipate as the hot plasma is advected outward and mixed with the cooler external material ($T_\text{e} < 10^{10} K$), leaving only weak residual structures near the eruption site.

The third row of Fig.~\ref{fig:vr_Te} shows the non-thermal electron fraction, $R/(1+R)$, where $R$ is given by Eq.~\eqref{ratioR}. Before and after the eruption, this fraction remains low across most of the disk, except in the highly magnetized region near the event horizon. At the eruption peak ($t = 11330\, t_\text{g}$), the non-thermal population is strongly enhanced within the expelled low-density region.
The fourth row shows the fractional vertical magnetic-field component, $\bar{B}_\theta/B$, which remains relatively small throughout the evolution. However, at the eruption peak, a localized patch at the periphery of the expelled flux ropes exhibits a pronounced enhancement of the vertical component. This behavior is consistent with the flux-expulsion scenario seen in GRMHD simulations \citep{Ripperda_2022}, in which reconnection-generated flux ropes are ejected outward and the vertical field is amplified in their surrounding sheath as field lines are compressed.
Within our field-aligned beaming eDF model, the spatial coincidence of enhanced non-thermal electrons and locally vertical magnetic fields temporarily provides a favorable geometry for anisotropic emission toward a near-axis observer, producing the discontinuous bright ``blobs'' superposed on the dimmed background. In the post-eruption stage, although the vertical-field fraction increases over a broader region, the non-thermal electron fraction has already declined significantly (as seen in the third row). Consequently, this late-time vertical field does not give rise to strong anisotropic non-thermal emission, and the images remain overall dim.

\begin{figure*}[htbp]
	\centering
\hspace*{-1cm}
	\includegraphics[width=0.77\textwidth]{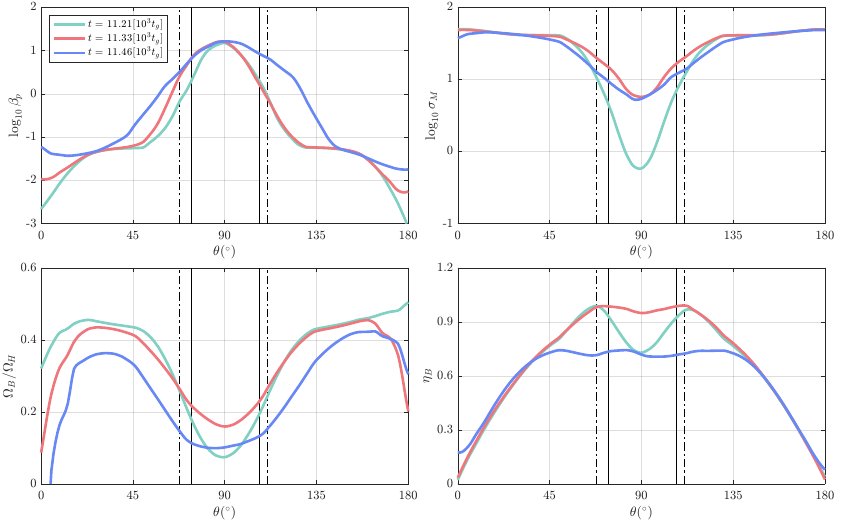}
	\caption{Variations of plasma-$\beta$ (\textbf{top left}), magnetization parameter $\sigma_\text{M}$ (\textbf{top right}), field-line angular velocity $\Omega_\text{B}$ (\textbf{bottom left}) and the field winding $\eta_\text{B}$ (\textbf{bottom right}) as functions of $\theta$, at $t = 11210\,t_\text{g}$, $11330\,t_\text{g}$ and $11460\,t_\text{g}$. For each cone of constant $\theta$, all quantities are averaged over $r \in (r_\text{h}, 10)$  and $\phi \in (0, 2\pi)$, where $r_\text{h} = M + \sqrt{M^2- a^2}$ is the horizon radius. 
In each panel, the solid and dash-dotted lines mark the time- and angle-averaged contours of $Be = 1.05$ and $\sigma_\text{M} = 20$, respectively.}
	\label{fig:thetaplot}
\end{figure*}

To further clarify the flow structure during the eruption episode, we plot in Fig.~\ref{fig:thetaplot}  the angular distributions of several key quantities at $t = 11210 \, t_\text{g}$, $11330 \, t_\text{g}$ and $11460 \, t_\text{g}$. Each quantity is averaged over $r \in (r_\text{h}, 10)$  and $\phi \in (0, 2\pi)$ on cones of constant $\theta$.
The top-left panel of Fig.~\ref{fig:thetaplot} shows the $\t$-dependence of the plasma-$\beta$.
Near the equatorial plane, the accretion flow is matter-dominated, with $\beta \simeq 10$, and the magnetic field lines are correspondingly more disordered (see Fig.~\ref{fig:rho}). 
In contrast, the jet region has $\beta < 10^{-1}$, indicating magnetic dominance. During the eruption, we observe a modest increase in $\beta$ across the disk region.
The top-right panel of Fig.~\ref{fig:thetaplot} shows the angular profile of the magnetization parameter $\sigma_\text{M}$. 
The jet remains strongly magnetized with $\sigma_\text{M} \gtrsim 20$.
Near the equatorial plane, $\sigma_\text{M}$ is initially below unity, but during the eruption it increases substantially, with values well above $1$, indicating the development of a more magnetically dominated zone within the main emission region.

The bottom panels in Fig.~\ref{fig:thetaplot} show the angular distributions of the field-line angular velocity and winding degree. The winding is quantified by the angle $\eta_\text{B}$
between the toroidal and poloidal magnetic-field components, while the field-line angular velocity $\Omega_\text{B}$ characterizes how the magnetic field couples to the plasma flow and black hole rotation under ideal MHD. The original definition of $\Omega_\text{B}$ takes $\Omega_\text{B} = F_{tr}/F_{r\phi} = F_{t\t}/F_{\t\phi}$ \citep{Blandford1977, thorne1982electrodynamics}. Here, we adopt an alternative formulation appropriate for GRMHD simulations \citep{McKinney2012}. Putting these together, we have 
\begin{equation}\label{eq:Brotation}
	\Omega_\text{B} = v^\phi - B^\phi \frac{\bar{v}_r \bar{B}_r + \bar{v}_\theta \bar{B}_\theta}{\bar{B}_r^2 + \bar{B}_\theta^2} \,, \quad \eta_\text{B} =  \arctan \left( \f{\bar{B}_\phi}{\sqrt{\bar{B}_r^2 + \bar{B}_\theta^2}} \right)\,.
\end{equation}
Because the $r$ and $\phi$ coordinates in Kerr–Schild spacetime are not orthogonal, we employ the ``quasi-orthogonal'' prescription \citep{McKinney2012} to evaluate the effective magnitudes of 3-vectors along each coordinate direction. For example, for the magnetic field $B^i$, we define $\bar{B}_i = \sqrt{g_{ii}} B^i$ as its magnitude in the corresponding direction.

The bottom-left panel of Fig.~\ref{fig:thetaplot} shows that in the jet region, $\Omega_\text{B}$ can reach $\sim 0.5 \Omega_\text{H}$, where $\Omega_\text{H} = a/(2 r_\text{h})$ is the black hole angular velocity, consistent with an efficient Blandford-Znajek process \citep{Blandford1977}.
In the disk region, $\Omega_\text{B}\simeq 0.1\Omega_\text{H} - 0.2 \Omega_\text{H}$, primarily reflecting the sub-Keplerian rotation of the accreting matter. 
As magnetic flux is released and disk material is expelled, field lines previously anchored to the horizon move into the disk region, causing $\Omega_\text{B}$ there to increase from $0.1 \Omega_\text{H}$ at $t = 11210 \, t_g$ to $0.2 \Omega_\text{H}$ at $t = 11330 \, t_g$.  
In the post-eruption stage, the growing turbulence further distorts the magnetic-field configuration, leading to a latitude-wide reduction of $\Omega_\text{B}$ by about $0.1 \Omega_\text{H}$.

The bottom-right panel of Fig.~\ref{fig:thetaplot} shows the angular distribution of the field-line winding $\eta_\text{B}$. 
Toward the polar axis, the magnetic field becomes increasingly poloidal, as is typical for relativistic jets \citep{CruzOsorio2022}, whereas in the disk region the field develops a more pronounced toroidal component.
Before the eruption, turbulence in the near-equatorial, matter-dominated region reduces the $\phi$-averaged value of $\eta_\text{B}$, as indicated by the green curve.
During the eruption, ordered field lines that thread the horizon are advected into the disk region, producing an increase in $\eta_\text{B}$ there.
After the eruption, the expansion of turbulence disrupts the magnetic-field structure and suppresses $\eta_\text{B}$ over a broader angular range, as indicated by the purple curve.

\section{Numerical Framework for GRRT}\label{app:coport}
Building upon the \textbf{Coport} framework \citep{Huang2024}, we have optimized the numerical scheme for the radiative transfer equation to more efficiently exploit the adaptive mesh refinement (AMR) grid data produced by the BHAC code. We refer to this improved numerical scheme as \textbf{Coport-2.0}, which is implemented using modern C++. \textbf{Coport-2.0} can perform approximately $500$ radiative transfer equation calculations per second on a single thread of modern multi-core CPUs. This section presents the details of the improved numerical method.
The photon trajectory is obtained by integrating the geodesic equations:
\begin{align}\label{eqn:geo1222}
	\frac{\mathrm{d} x^\mu}{\mathrm{d} \lambda}=k^\mu\,,\quad
	\frac{\mathrm{d} k^\mu}{\mathrm{d} \lambda}=-\Gamma^\mu_{\ \nu\rho}k^\nu k^\rho\,, \quad
	\frac{\mathrm{d} f^\mu}{\mathrm{d} \lambda}=-\Gamma^\mu_{\  \nu\rho}k^\nu f^\rho\,,
\end{align}
where $x^\mu$ and $k^\mu$ denote the photon's position and four-momentum, respectively; $\Gamma^\mu_{\ \nu\rho}$ are the Christoffel symbols; and $\lambda$ is the affine parameter. The vector $f^\mu$ defines a polarization basis orthogonal to $k^\mu$ and is parallel-transported along the geodesic.
The evolution of the Lorentz-invariant Stokes vector takes $\vec{\Mc{S}}=\left({\Mc{I},\Mc{Q},\Mc{U},\Mc{V}}\right)$, where $\Mc{I}$ is the covariant total intensity, $\Mc{Q}, \Mc{U}$ are covariant linear polarization degrees, and $\Mc{V}$ denotes the circular polarization degree. 
The evolution of $\vec{\Mc{S}}$ is governed by the radiative transfer equation \citep{Broderick:2003fc, Shcherbakov:2010kh}:
\begin{align}
	\dfrac{1}{\Mc{C}}
	\dfrac{\mathrm{d}}{\mathrm{d}\lambda}
	\begin{pmatrix}
		\Mc{I} \\ \Mc{Q}\\ \Mc{U}\\ \Mc{V}
	\end{pmatrix}
	=
	R(\chi)\begin{pmatrix}
		j_{I} \\ j_{Q}\\ 0 \\ j_V
	\end{pmatrix}-
	\left[R(\chi)
	\begin{pmatrix}
		\a_I & \a_Q & 0 & \a_V\\
		\a_Q&\a_I&\rho_V&0\\
		0&-\rho_V&\a_I&\rho_Q\\
		\a_V&0&-\rho_Q&\a_I
	\end{pmatrix}
	R(-\chi)\right]
	\begin{pmatrix}
		\Mc{I} \\ \Mc{Q}\\ \Mc{U}\\ \Mc{V}
	\end{pmatrix}\,,
	\label{eqn:fszy1212}
\end{align}
where $\mathcal{C}=GM/c^2\nu_0$ is the dimensionless conversion factor; $\nu_0$ denotes the photon frequency observed at infinity; $j_i$, $\a_i$, $\rho_i$ are the (invariant) emissivities, absorptivities, and Faraday coefficients. The rotation matrix $R(\chi) = R^{-1}(-\chi)$ is parameterized by the angle $\chi$ between $f^{\mu}$ and the local plasma magnetic field $b^{\mu}$. It can be expressed in a covariant form:
\begin{align}
	\chi=\text{sign}&(\epsilon_{\mu \nu \rho \sigma  }u^\mu f^\nu b^\rho k^\sigma)\times
	\arccos\left(\dfrac{P^{\mu \nu}f_{\mu }b_{\nu}}{\sqrt{ (P^{\mu \nu}f_{\mu}f_{\nu} )(P^{\alpha \beta}b_{\alpha }b_{\beta} )}}\right) \,,
	\label{eqn:jdjs}
\end{align}
where $u^\mu$ is the fluid 4-velocity, $P^{\mu \nu}$ the induced metric on the subspace orthogonal to both $u^\mu$ and $k^\mu$:
\begin{align}	
	P^{\mu \nu}=g^{\mu \nu}+u^\mu u^\nu-e_{k}^\mu e_{k}^\nu\,,
	\qquad e_{k}^\mu= \frac{k^\mu}{\nu}-u^\mu\,,
\end{align}
where $\nu = -k_\mu u^\mu$ represents the frequency of photons as observed by the co-moving fluid frame.
\par 
Given the different convergence radii and stiffness characteristics of Eqs.\eqref{eqn:geo1222}, \eqref{eqn:fszy1212}, distinct numerical schemes are employed for each equation to improve both accuracy and convergence behavior.
In particular, Eq.\eqref{eqn:geo1222} is integrated using an adaptive Runge–Kutta method, where the input step size $\Delta \lambda_{\text{grid}}$ is set by the AMR grid cell length $\Delta x^i$ and the spatial components of the photon four-velocity $k^i$:
\begin{align}
	\Delta \lambda_{\text{grid}}=\dfrac{1}{n}\text{min}\left\{\dfrac{\Delta x^i}{k^i}\right\}\,,\qquad i\in\{1,2,3\}\,,
\end{align}
Here, $n$ specifies the number of integration steps within each grid cell. The adaptive scheme subsequently refines $\Delta \lambda_{\text{grid}}$ to satisfy the prescribed error tolerance, yielding an effective step size $\Delta \lambda$ ($\Delta \lambda\leq \Delta \lambda_{\text{grid}}$). Once $\Delta \lambda$ is determined, the formal solution introduced in \citep{degl1985solution} is applied to integrate Eq.\eqref{eqn:fszy1212}, thereby ensuring numerical stability and consistency of the results.
\par
Before solving the radiative transfer equation Eq.~\eqref{eqn:fszy1212}, we pre-process the variables $\mathbf{P} = \{\rho, e, \tilde{u}^i, \tilde{B}^i\}$ provided by \textbf{BHAC}~\citep{Porth2017}. Here, $\rho$ denotes the rest-mass density, $e$ the internal energy density of the gas, $\tilde{u}^i = \Gamma v^i$ the spatial components of the four-velocity in the Eulerian frame, $\Gamma = 1/\sqrt{1 - v^2}=\sqrt{1+\tilde{u}^2}$ the Lorentz factor, and $\tilde{B}^i$ the magnetic field measured by an Eulerian observer.
For a photon located at position $\vec{X}$, we identify the eight cell vertices $\vec{x}_i$ surrounding the point and compute the linear interpolation of the primitive variables weighted by the corresponding volume fractions $\mu_i$:
\begin{align}
	\bar{\mathbf{P}}=\sum_{i=1}^{8}\mu_i \mathbf{P}_i\,,
	\qquad  
	\mu_i=\prod_{j=1}^3 \left(1-\left|\dfrac{x_i^j-X^j}{\Delta x^j}\right|\,\right)\,.
\end{align}
Using the interpolated primitive variables $\bar{\mathbf{P}}$, the fluid four-velocity $u^{\mu}$ and the magnetic four-vector $b^{\mu}$ are computed as follows:
\begin{equation}
	u^{\mu} = \Gamma \left( \frac{1}{\alpha}, \dfrac{\tilde{u}^i}{\Gamma} - \frac{\beta^i}{\alpha} \right)
	\,,\qquad
	b^\mu=\left(
	\frac{\tilde{B}_j \tilde{u}^j}{\alpha},
	\frac{\tilde{B}^i + (\tilde{B}_j \tilde{u}^j) u^{i}}{\Gamma}
	\right)\,,
\end{equation}
where $\alpha$ and $\beta^i$ denote the lapse function and the shift vector in the 3+1 decomposition of spacetime, respectively.
At the observer's location, the Stokes vector $\vec{\Mc{S}}$ must be rotated to align with the image plane’s coordinate frame. 
The rotation angle $\chi_o$ is calculated by producting $f^{\mu}$ with the image-plane $y$-axis basis $(\propto -\partial_\theta^\mu)$, which yields $\chi_o = \text{sign}(f^{\t}) \arccos{\left(f^{\phi}/f\right)}$. The observed Stokes vector, $\vec{\Mc{S}}_o=\left({\Mc{I}_o,\Mc{Q}_o,\Mc{U}_o,\Mc{V}_o}\right)$, is then obtained through
\begin{align}
	\Mc{I}_o=\Mc{I}\,,\quad
	\Mc{Q}_o=\Mc{Q}\cos\chi_o-\Mc{U}\sin\chi_o\,,\quad
	\Mc{U}_o=\Mc{Q}\sin\chi_o+\Mc{U}\cos\chi_o\,,\quad
	\Mc{V}_o=\Mc{V}\,.
\end{align}

\section{Supplementary Plots of the Images}\label{app:imaging}

\begin{figure*}[htbp]
	\centering
	\includegraphics[width=0.7\textwidth]{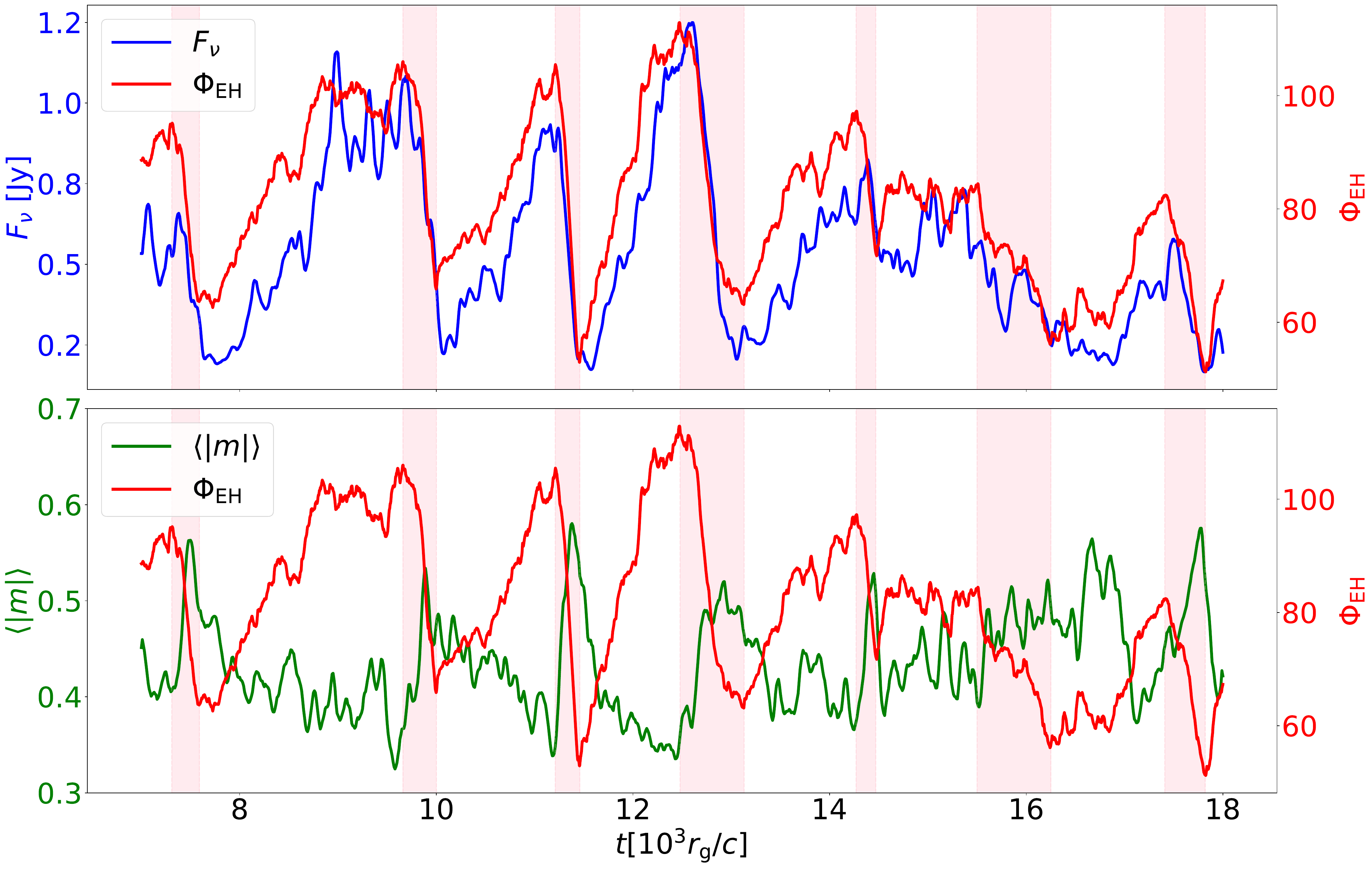}\\~\\
    \hspace*{-1cm}\includegraphics[width=4.75in]{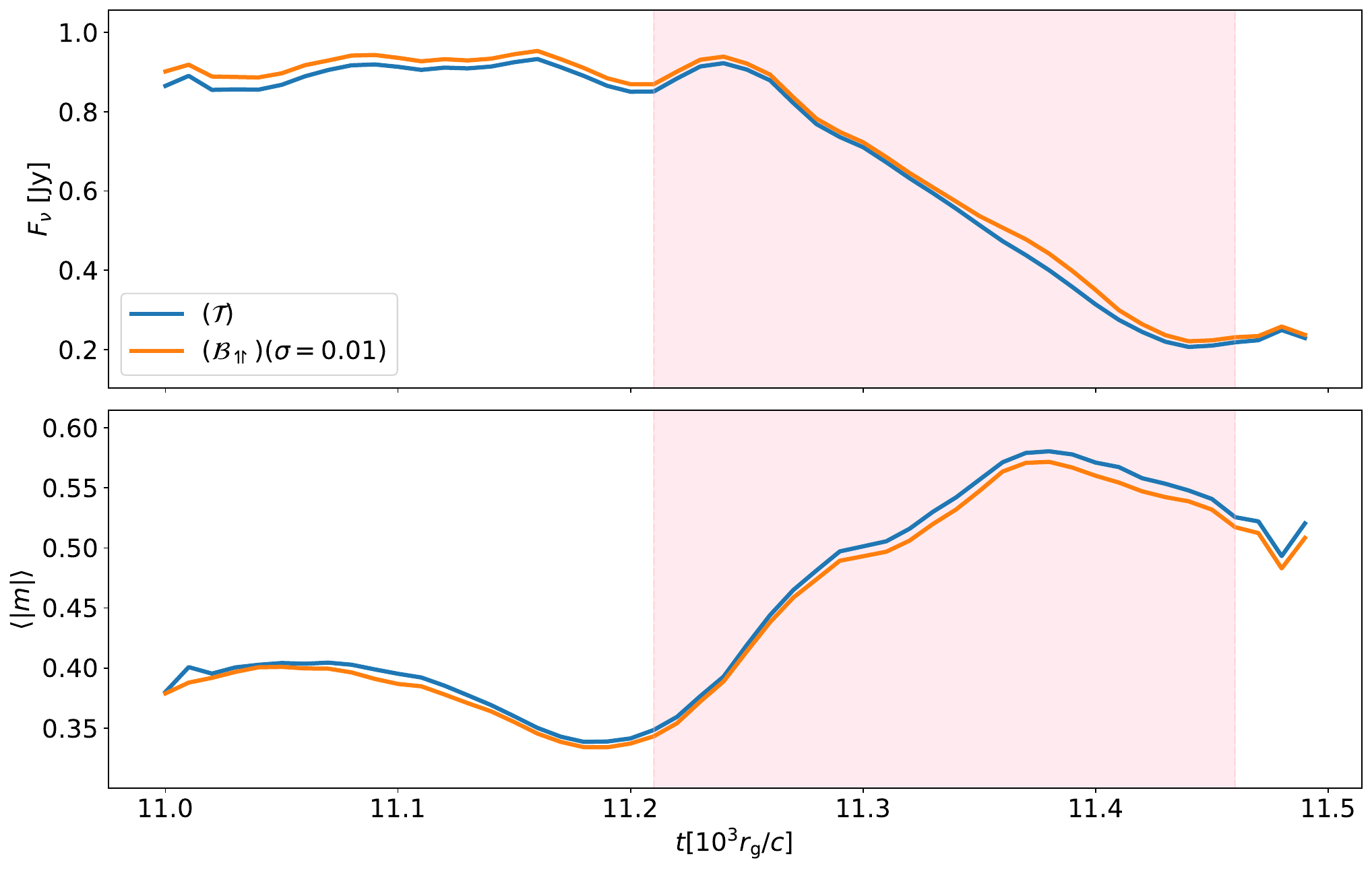}
	\caption{\textbf{Top:} Time evolution of the intrinsic flux and linear polarization fraction under thermal electron distribution for different eruption events.
	The variations are similar across each phase: the flux decreases with the rapid decline of magnetic flux, while the linear polarization degree increases as the magnetic flux decreases.
\textbf{Bottom:} Comparison of the intrinsic flux and linear polarization fraction between the thermal model and a hybrid thermal+beam model with $\sigma = 0.01$ for the third eruption event.}
	\label{fig:flux_diffevent}
\end{figure*}

%


\clearpage

\bibliographystyle{aasjournal}
\bibliography{main}

\end{document}